\documentclass[12pt]{academia}

\usepackage{graphicx}

\usepackage{amssymb}

\begin{document}

\title{THE IMPACT OF GIULIO RACAH ON CRYSTAL- AND LIGAND-FIELD THEORIES}
\author{MAURICE R. KIBLER
\\
\small{Universit\'e de Lyon, 37 rue du Repos, F--69361 Lyon, France} \\
\small{Universit\'e Claude Bernard, 43 Bd du 11 Novembre 1918, F--69622 Villeurbanne, France} \\
\small{CNRS~/~IN2P3, IPNL, 4 rue Enrico Fermi, F--69622 Villeurbanne, France}}
\date{\today}

\maketitle


    \vspace*{-10.4cm} \noindent{\sl Paper based on an invited talk given at the ``International 
    Conference in Commemoration of the Centenary of the Birth of G. Racah (1909-1965)'' 
    (Zaragoza, Spain, 22-24 February 2010).} 
    \vspace*{9cm}

\setcounter{page}{1}

\label{BartolomeP}

\begin{abstract}

This paper focuses on the impact of Racah on crystal- and ligand-field theories, two branches 
of molecular physics and condensed matter physics (dealing with ions embedded in aggregates of 
finite symmetry). The role of Racah and some of his students in developing a symmetry-adapted 
weak-field model for crystal-field theory is examined. Then, we discuss the extension of this 
model to a generalized symmetry-adapted weak-field model for ligand-field theory. Symmetry 
considerations via the use of the Wigner-Racah algebra for chains of type $SU(2) \supset G$ 
is essential for these weak-field models. Therefore, the basic ingredients for the Wigner-Racah 
algebra of a finite or compact group are reviewed with a special attention paid to the $SU(2)$ 
group in a $SU(2) \supset G$ basis. Finally, as an unexpected application of nonstandard $SU(2)$ 
bases, it is shown how $SU(2)$ bases adapted to the cyclic group allow to build bases of 
relevance in quantum information.

\end{abstract}

\section{Introduction} \label{sec_intro}

The legacy of Giulio Racah (Firenze, 1909-1965) stems mainly from his four papers {\em Theory of complex spectra} 
published between 1942 and 1949 \cite{RI, RII, RIII, RIV}, his notes on group-theoretical methods in spectroscopy 
based on lectures given at the Institute for Advanced Study in Princeton in 1951 \cite{IASP, Springer}, and his 
book on irreducible tensorial sets written in collaboration with his cousin Ugo Fano \cite{FanoRacah}. 

It was the purpose of the first paper of his famous series \cite{RI, RII, RIII, RIV} to substitute to the 
numerical methods of Slater, Condon and Shortley general methods more conformable to the Dirac representation 
of state vectors. The main achievements realized in \cite{RI, RII, RIII, RIV, IASP, Springer, FanoRacah} 
deal with irreducible tensor methods, Wigner-Racah algebra (or Racah-Wigner algebra or Racah algebra, 
a concept to be precisely defined in Section 3) and group-theoretical methods involving chains 
of Lie groups. More precisely, let us mention the following important contributions. 
\begin{itemize}
	\item The development of the algebra of coupling and recoupling coefficients for the $SU(2)$ group in a $SU(2) \supset U(1)$ 
  basis, with introduction of the $V$ and $\overline{V}$ {\em functions} (the $\overline{V}$ symbol is identical 
  to the 3--$jm$ Wigner symbol up to a permutation of its columns) and of the $W$, $\overline{W}$ and $X$ {\em functions} 
  (the $\overline{W}$ and $X$ symbols are identical to the 6--$j$ and 9--$j$ Wigner symbols, respectively). 
	\item The introduction of the concept of a $SU(2)$ irreducible {\em tensor operator} that generalizes the notion
	of a vector operator and the generalization to tensor operators of the Wigner-Eckart theorem for vector operators. 
	\item The introduction of the notion of a {\em unit tensor operator}, the matrix elements of which in a $SU(2) \supset U(1)$ 
  basis are nothing but Clebsch-Gordan coefficients (up to a multiplicative factor), with the advantage that any tensor operator 
  is proportional to a unit tensor operator.  
	\item The introduction of the concept of {\em seniority} which is related to the state labeling problem.
	\item The development of the notion of {\em coefficients of fractional parentage},	previously introduced 
	by Goudsmit and Bacher, which make it possible to develop a $n$--particle wavefunction in terms of $(n-1)$--particle 
	wavefunctions.
	\item The introduction of {\em chains of Lie groups}, involving both invariance and classification groups, 
	for characterizing state vectors and interactions involved in 
	spectroscopic problems. To implement the use of chains of groups, Racah introduced a {\em factorization lemma} and 
	developed the notion of a {\em complete set of commuting operators} (involving Cartan operators, invariant or 
	Casimir operators and labeling operators) in a group-theoretical context.
\end{itemize}

The series of seminal works \cite{RI, RII, RIII, RIV, IASP, 
Springer, FanoRacah} opened the way for many 
applications by Racah himself, his students and a large part of the community of scientists working in atomic and nuclear 
spectroscopy (see the list of Racah's publications in \cite{Talmi}). In particular, the methods of Racah were popularized 
by Judd \cite{Juddbook}, Wybourne \cite{Wybournebook}, and Condon and Odaba\c{s}i \cite{Condon} in atomic physics and by 
de-Shalit and Talmi \cite{ShalitTalmi} in nuclear physics (see also \cite{Iachello, Isacker1, Isackerentre1et2, Isacker2} for recent developments in 
nuclear and molecular physics). The basic concepts introduced 
and/or developed by Racah in his pioneer works  
were also of considerable importance in molecular and condensed matter physics. More specifically, these 
works stimulated an enormous quantity of developments in vibration-rotational spectroscopy of molecules 
and in crystal- and ligand-field theories as will be shown below. 

We shall be concerned here with the impact of Racah on crystal- and ligand-field theories, two theories which deal with 
optical, magnetic and thermal properties of ions embedded in molecular, condensed matter or biological surroundings. Racah 
never published papers about these theories. However, he was interested in molecular physics as shown by the fact that 
he published in 1943 a short note on the structure of the Mo(CN)$_8^{4-}$ complex ion \cite{Racah-molecular}. His interest 
for molecular physics and the physics of ions in crystals was reinforced and stimulated by a seminar given by his colleague 
Willy Low in the Department of Physics of the Hebrew University of Jerusalem in 1956 \cite{Low, L60, L61}. The seminar was 
devoted to the role of crystalline fields on the optical spectra of transition-metal ions (like Ni$^{2+}$ and Co$^{2+}$) 
in crystals. Racah became very much interested in this research subject and decided to guide students 
in this direction. His idea was to combine his irreducible tensor methods with the group-theoretical methods largely used in 
crystal-field theory (but limited in those times to a qualitative explanation of the level splitting for a given ion embedded 
in a finite symmetry surrounding). Along this line, Racah and Low directed two graduate students, Schoenfeld who studied the 
case of the $d^2$ and $d^3$ configurations in cubic symmetry \cite{Schoenfeld} and Rosengarten who dealt with the case of 
$d^4$ and $d^5$ configurations in the same symmetry \cite{LowRosengarten}. Then, Racah asked another student, Flato, to work out 
the more involved case of the $d^2$ and $d^3$ configurations in trigonal and tetragonal symmetries \cite{Flato1}. Five years 
after having completed his thesis, the material contained in Flato's thesis was still 
of such an interest that he was asked to publish it (for the main part) \cite{Flato2} 
(see also \cite{MScthesisFlato}). Research in that direction continued with a 
general formalism and a symmetry-adapted weak-field model developed by the present 
author in his thesis prepared under the guidance of Flato \cite{JMS68,IJQC69,CRAS}. 

It is one of the aims of the present review to show how Racah directly and indirectly contributed to the penetration 
in crystal- and ligand-field theories of the tools he originally developed for atomic and nuclear spectroscopy. 
Another aim of this article is to show how the Wigner-Racah algebra for a group of molecular or crystallographic interest can 
be deduced from the one of $SU(2)$ in a nonstandard basis. To a large extent, this paper constitutes a brief review of the methods 
and models used in crystal- and ligand-field theories as well as a pedestrian presentation of the Wigner-Racah algebra for a 
chain of groups involving finite and/or compact groups. As an application of the $SU(2) \supset G$ chain, where $G$ is a 
cyclic group, a brief contact is established with quantum information, a field of considerable interest in the present days.    

The material in this paper is organized as follows. Section 2 deals with crystal- and ligand-field theories. The basic ingredients 
for the Wigner-Racah algebra of a finite or compact group together with some illustrative examples are given in Section 3. Section 
4 is devoted to a short incursion in quantum information via the use of specific chains of type $SU(2) \supset G$. 

Most of the notations are standard. The star denotes complex conjugaison, $\delta_{a   b}$ the Kronecker 
delta symbol of $a$ and $b$, 
and $A^{\dagger}$ the adjoint of the operator $A$.
We use a notation of type $\vert \psi )$ (as in Racah's papers), or $\vert \psi \rangle$ 
(as in quantum information), for a vector in an Hilbert space and we denote 
$\langle \phi \vert \psi \rangle$ and $\vert \phi \rangle \langle \psi \vert$ respectively the inner and 
outer products of the vectors $\vert \psi \rangle$ and $\vert \phi \rangle$. 
Specific notations on group theory shall be introduced later.

\section{Crystal- and ligand-field theories}

\subsection{Generalities}

Crystal- and ligand-field theories deal with the description and interpretation of electronic and 
magnetic properties (optical spectra, electron paramagnetic resonance spectra, photo-electron 
spectra, etc.) of a partly-filled shell ion in a molecular, condensed matter or biological environment.  
Crystal-field theory (based on the use of atomic orbitals) goes back to the end of the twenties 
with the seminal work by Bethe \cite{Bethe} and was applied to the spectroscopy of ions in solids 
in the early days of quantum mechanics. 
It is only in the fiftees that ligand-field theory (based 
on the use of molecular orbitals) was the object of numerous studies.
In modern parlance, 
crystal- and ligand-field theories are special cases of the theory of level splitting. 

As a typical example, let us consider the case of a ruby crystal. It consists of corindon (Al$_2$O$_3$) 
doped with trivalent chromium  ions (Cr$^{3+}$) in substitution with 
           trivalent aluminium ions (Al$^{3+}$). The electrons of 
each Cr$^{3+}$ ion are thus subjected to inhomogeneous electric fields arising from the ligands or coordinats 
constituted by the oxygen atoms. These electric (or crystalline) fields yield a level splitting of the 
energy levels of the Cr$^{3+}$ ion. One-photon transitions in the visible between the split levels are 
responsible for the nice pink to blood-red color of ruby. 

The distinction between crystal-field theory and ligand-field theory is as follows. In crystal-field theory 
one uses atomic orbitals for the central partly-filled shell ion (the Cr$^{3+}$ ion in our example) whereas 
in ligand-field theory one considers molecular orbitals made of linear combinations of atomic orbitals of 
the central ion and of the ligands or coordinats (the O$^{2-}$ ions in our example). 

\subsection{The Hamiltonian}

We shall consider the common case of an ion with a $\ell^N$ atomic configuration ($N$ equivalent electrons on a 
$n \ell$ shell outside of a set of closed shells). The $\ell = 2$ case corresponds to transition metal-ions and 
the $\ell = 3$ case to rare earth and actinide ions. In first approximation, the perturbation Hamiltonian 
${\cal H}$ for such an ion embedded in a crystalline field reads
      \begin{eqnarray}
{\cal H} := {\cal H}_C + {\cal H}_{so} + {\cal H}_{cf}
      \label{hamilonian}
      \end{eqnarray}
where ${\cal H}_C$ stands for the two-body Coulomb interaction between the $N$ electrons, ${\cal H}_{so}$ the 
one-body spin-orbit interaction for the $N$ electrons and ${\cal H}_{cf}$ the one-body interaction between the 
$N$ electrons and the environment of the central ion. Obviously, ${\cal H}_C$ and ${\cal H}_{so}$ are rotationally 
invariant and ${\cal H}_{cf}$ is invariant under the point symmetry group $G$ of the ion and its surrounding. Therefore, 
the $O(3) \supset G$ chain of groups naturally plays an important role in the description of the ion in its environment 
(the three-dimensional orthogonal group $O(3)$ is isomorphic with the three-dimensional rotation group). When $G$ contains 
only rotations, it is sufficient to consider the $SO(3) \supset G$ chain ($SO(3)$ is the three-dimensional special orthogonal 
group) or the $SU(2) \supset G^*$ chain, where $SU(2)$ and $G^*$ are the spinor groups (double groups in the terminology of Bethe) 
of $SO(3) \sim SU(2)/Z_2$ and $G \sim G^*/Z_2$, respectively. We can thus understand the importance of both continuous and finite 
groups in crystal- and ligand-field theories.  

In view of the various terms in ${\cal H}$, we can have several families of models. The situations
      \begin{eqnarray}
{\cal H}_C > {\cal H}_{so} > {\cal H}_{cf}
      \label{weak}
      \end{eqnarray}
and
      \begin{eqnarray}
{\cal H}_{cf} > {\cal H}_C > {\cal H}_{so}
      \label{strong}
      \end{eqnarray}
correspond to the so-called weak-field model and the strong-field model, respectively. The strong-field model was mainly 
developed in the fifties by Tanabe, Sugano and Kamimura in Japan \cite{TS541, TS542, TS56, TK581, ST582, stk70} and by 
Griffith in England \cite{GTFS1, GTFS2, GTFS3, GMP1, GMP2, GMP3, GMP4, GMP5, Livre1, griffith62}, and later by Tang 
Au-chin and his collaborators in China \cite{chinois1, chinois2, chinois3, chinois4, chinois5} as well as by Smirnov and his 
collaborators in the former USSR \cite{Smirnov1, Smirnov2, Smirnov3, Smirnov4, Smirnov5}. The weak-field model, although 
worked out in the early days of crystal-field theory, was systematically developed from the sixties. In particular, a 
symmetry-adapted version of the weak-field model was introduced, as we mentioned in the introduction, following a suggestion of Racah 
by two of his students, Schoenfeld \cite{Schoenfeld} and Flato \cite{Flato1, Flato2}. It was further developed by the present author 
and some of his collaborators \cite{JMS68, IJQC69, IJQC3, IJQC4, IJQC5, CCA6} (see also \cite{chinois1, chinois2, chinois3, chinois4, chinois5} 
and \cite{Smirnov1, Smirnov2, Smirnov3, Smirnov4, Smirnov5}). In crystal-field theory, the weak- and strong-field models 
are {\em a priori} equivalent if the matrix of ${\cal H}$ is set up on the 
      \begin{eqnarray}
C_{4 \ell + 2}^N := \frac{(4 \ell + 2)!}{(4 \ell + 2 - N)! N!} 
      \label{dimension}
      \end{eqnarray}
state vectors of the $\ell^N$ configuration. Nevertheless, the implementations of the two models are quite different as it 
will be shown below. As an illustration, we shall now discuss in turn the two models (strong- and weak-field models) in 
the special case of $d^N$ ions in cubic symmetry. 

\subsection{Strong-field models}
 
It is difficult to describe the strong-field model in the general case of $\ell^N$ in $G$. Hence, we consider the case of 
a $d^N$ ion ($\ell = 2$) in octahedral symmetry ($G = O$). The restriction $SO(3) \to O$ yields the following decomposition
      \begin{eqnarray}
2 = E \oplus T_2
      \label{restriction O3-O}
      \end{eqnarray}
of the irreducible representation class (IRC) of $SO(3)$ associated with $\ell = 2$ into a direct sum of the IRCs $E$ and $T_2$ 
of finite group $O$. As a consequence, there is a splitting level: the five degenerate $d$ orbitals give rise to a $E$ doublet  
(with two degenerate orbitals $e$) and a $T_2$ triplet (with three degenerate orbitals $t_2$). The $e$ and $t_2$ orbitals can 
be considered as symmetry-adapted atomic orbitals (in crystal-field theory) or as molecular orbitals (in ligand-field 
theory). The distribution of the $N$ electrons on the $t_2$ and $e$ orbitals, according to the Pauli exclusion principle, 
yields (molecular) configurations of type 
$t_2^x e^{N-x}$. Then, we can form (molecular) terms $t_2^x(S_1 \Gamma_1) e^{N-x}(S_2 \Gamma_2)$, where $S_1$ and $S_2$ 
are the total spins for the $x$ and $N-x$ electrons on the $t_2$ and $e$ orbitals, respectively. Furthermore, $\Gamma_1$ 
(contained in $T_2^{\otimes x}$) and $\Gamma_2$ (contained in $E^{\otimes (N-x)}$) denote the IRCs characterizing the 
orbital parts of the $t_2$ and $e$ electrons. The next step is to couple $S_1$ with $S_2$ to get the total spin $S$ 
(contained in $S_1 \otimes S_2$) and $\Gamma_1$ with $\Gamma_2$ to obtain $\Gamma$ 
(contained in $\Gamma_1 \otimes \Gamma_2$). This leads to (molecular) states 
$t_2^x(S_1 \Gamma_1) e^{N-x}(S_2 \Gamma_2) S \Gamma$. Finally, the coupling of $S$ (decomposed into IRCs of $O^*$) with $\Gamma$ 
gives the total IRC $\Gamma_T$ (an internal branching multiplicity label $b$ is necessary when $\Gamma_T$ occurs several times 
in the reduction of $S \otimes \Gamma$). As a result, we get state vectors of type 
      \begin{eqnarray}
| t_2^x(S_1 \Gamma_1) e^{N-x}(S_2 \Gamma_2) S \Gamma b \Gamma_T \gamma_T)
      \label{strong-field vectors}
      \end{eqnarray}
which are expressed (via complicated formulas) in terms of one-electron state vectors by means of coupling coefficients 
and coefficients of fractional parentage. Note that the label $\gamma_T$ in (\ref{strong-field vectors}) is necessary 
when the dimension of $\Gamma_T$ is greater than 1. 

The calculation of the matrix elements of ${\cal H}_{cf}$ in the strong-field basis (\ref{strong-field vectors}) is 
elementary. However, this is not the case for ${\cal H}_C + {\cal H}_{so}$. The construction of the matrix of 
${\cal H}_C + {\cal H}_{so}$ on $C^{N}_{4 \ell +2}$ state vectors (\ref{strong-field vectors}) requires 
the knowledge of coupling and recoupling coefficients for both $SU(2)$ and $G^*$ as well as coefficients of 
fractional parentage for the configurations $t_2^x e^{N-x}$. 

From the practical point of view, the just described strong-field approach leads to: 
\begin{itemize}
	\item a five-parameter model in a crystal-field framework where the $t_2$ and $e$ orbitals are atomic orbitals, 
called ordinary strong-field model, with 3 parameters for ${\cal H}_C$ ($F_0$, $F_2$ and $F_4$ of Slater or $A$, $B$ 
and $C$ of Racah, see the appendix), 1 parameter for ${\cal H}_{so}$ ($\zeta_{n d}$) and 1 parameter for ${\cal H}_{cf}$ ($10 Dq$)
  \item a fourteen-parameter model in a ligand-field framework where the $t_2$ and $e$ orbitals are molecular 
orbitals, called generalized strong-field model, with 10 parameters for ${\cal H}_C$, 2 parameters for ${\cal H}_{so}$ 
and 2 parameters for ${\cal H}_{cf}$. 
\end{itemize}

The strong-field models present several drawbacks. The case of $d^N$ in $O$ is difficult to extend to the case 
of $\ell^N$ in $G$: replacing $O$ by $G$ and/or $d^N$ by $\ell^N$ requires that the calculation for ${\cal H}_C$ 
and ${\cal H}_{so}$, which involves complicated Wigner-Racah algebra developments for the $G$ or $G^*$ group  
with several phase problems, have to be done again. This kind of difficulty does not appear in a weak-field 
approach as shown below. 

\subsection{Weak-field models}
In the case of $\ell^N$ in $G$, we may think to use atomic state vectors of type $| n \ell^N \alpha SLJM )$. However, 
such state vectors, adapted to the $SU(2) \supset U(1)$ chain, are not generally adapted to the $G^*$ symmetry group. The 
idea of Racah was to use linear combinations of the vectors $| n \ell^N \alpha SLJM )$ transforming as IRCs of $G^*$ and 
to employ his methods for calculating the energy matrix of ${\cal H}$. Therefore, the matrices for ${\cal H}_C$ and 
${\cal H}_{so}$, in a $SU(2) \supset G^*$ symmetry-adapted basis, are the same as the ones of 
atomic spectroscopy (already calculated by Racah or easily calculable from Racah's methods) and the matrix of ${\cal H}_{cf}$ 
depends on reduced matrix elements of one-electron Racah unit tensor operators and $SU(2) \supset G^*$ symmetry-adapted Clebsch-Gordan 
coefficients. Thus, the implementation of the symmetry-adapted weak-field model is easier than the one of the 
ordinary strong-field model. Following Racah's idea, Schoenfeld and Flato calculated the matrix of ${\cal H}$ 
for the $d^2$ and $d^3$ configurations in cubic symmetry \cite{Schoenfeld} and in tetragonal and trigonal symmetries 
\cite{Flato1,Flato2}. Later, Low and Rosengarten dealt with the case of the $d^5$ configuration in cubic symmetry in 
connection with the optical spectra of Mn$_{2}^{2+}$ and Fe$^{3+}$ ions in crystalline fields \cite{LowRosengarten}. 

The Wigner-Racah algebra for the $SU(2)$ group in a $SU(2) \supset G^*$ symmetry-adapted basis of interest for the 
symmetry-adapted weak-field model was developed by the present author \cite{JMS68,IJQC69,CRAS,kibler79} and 
further considered by several authors \cite{Ellzey, Kremer, Rudra, Lulek, ButlerReid, Golding, Blaszak, Raynal, Del Duca}. The main 
ingredients of the resulting symmetry-adapted weak-field model for $\ell^N$ in $G$ can be summed up as follows. 

The symmetry-adapted weak-field state vectors are of type
      \begin{eqnarray}
| n \ell^N \alpha SLJ a \Gamma \gamma ) := \sum_{M=-J}^{J} | n \ell^N \alpha SLJM ) ( JM | J a \Gamma \gamma )
      \label{weak-field vectors}
      \end{eqnarray}
where $\Gamma$ is an IRC of $G^*$, $a$ a branching multiplicity label to be used when the $(J)$ IRC of $SU(2)$, 
associated with the $J$ quantum number, contains $\Gamma$ 
several times and $\gamma$ a multiplicity label to be used when the dimension of the $\Gamma$ IRC is greater than 1. In 
(\ref{weak-field vectors}), the $( JM | J a \Gamma \gamma )$ reduction coefficients are elements of a unitary matrix 
which reduces the representation matrix associated with the $(J)$ IRC of $SU(2)$ into a direct sum of representation matrices 
of $G^*$. They have to be distinguished from the reduction coefficients obtained from the diagonalization of an operator 
invariant under the $G$ group \cite{MoretJMS, Dijon2, Dijon3, Dijon4, PW1, PW2}. The $( JM | J a \Gamma \gamma )$ reduction 
coefficients are chosen in such a way that the set
      \begin{eqnarray}
\{ | n \ell^N \alpha SLJ a \Gamma \gamma ) : \gamma \ {\rm ranging} \}
      \label{set transforming as Gamma}
      \end{eqnarray}
spans a representation matrix associated with $\Gamma$ independent of the atomic quantum numbers and that the values of the 
corresponding coupling coefficients (the $f$ coefficients below) are square roots of rational numbers. Then, the matrices for 
${\cal H}_C$ and ${\cal H}_{so}$ follow from 
      \begin{eqnarray}
 ( n \ell^N \alpha SLJ a \Gamma \gamma | {\cal H}_C | n \ell^N \alpha' S' L' J' a' \Gamma' \gamma' ) = 
 \delta_{S S'} \delta_{L L'} \delta_{J J'} \delta_{a a'} \delta_{\Gamma \Gamma'} \delta_{\gamma \gamma'} \nonumber \\
 \times \Delta(S, L, J) ( n \ell^N \alpha S L M_S M_L| {\cal H}_C | n \ell^N \alpha' S L M_S M_L )
      \label{H Coulomb}
      \end{eqnarray}
and
      \begin{eqnarray}
( n \ell^N \alpha SLJ a \Gamma \gamma | {\cal H}_{so} | n \ell^N \alpha' S' L' J' a' \Gamma' \gamma' ) =  
\delta_{J J'} \delta_{a a'} \delta_{\Gamma \Gamma'} \delta_{\gamma \gamma'} \nonumber \\
 \times ( n \ell^N \alpha S L J M | {\cal H}_{so} | n \ell^N \alpha' S' L' J M )
      \label{H spin-orbit}
      \end{eqnarray}
where $\Delta(S, L, J)$ is 1 if $S$, $L$ and $J$ satisfy the triangular condition and 0 otherwise ; in (\ref{H Coulomb}) 
and (\ref{H spin-orbit}), the matrix elements in the right-hand 
sides are independent of the magnetic quantum numbers $M_S , M_L$ 
and $M$, respectively. Clearly, the energy matrices for ${\cal H}_C$ and ${\cal H}_{so}$ do not depend on the 
$G$ group and are easily builded from the works of Racah (the matrix elements in the 
right-hand sides of (\ref{H Coulomb}) and (\ref{H spin-orbit}) are known for the $p^N$, 
$d^N$ and $f^N$ configurations \cite{NielsonKoster} or easily calculable from 
computer programs). On the other hand, the matrix of ${\cal H}_{cf}$ can be readily set 
up by making use of the development
      \begin{eqnarray}
{\cal H}_{cf} = \sum_{k a_0} D[k a_0] U^{(k)}_{a_0 \Gamma_0 \gamma_0}
      \label{symmetry-adapted form of Hcf}
      \end{eqnarray}
where $U^{(k)}_{a_0 \Gamma_0 \gamma_0}$ is a component of a Racah unit tensor operator ${\bf U}^{k}$ invariant under $G$ (i.e., 
transforming as the $\Gamma_0$ identity IRC of $G$). In (\ref{symmetry-adapted form of Hcf}), $D[k a_0]$ are crystal-field 
parameters connected to the $B^k_q$ parameters (in Wybourne's normalization \cite{Wybournebook,RajnakWyb}) via
      \begin{eqnarray}
D[k a_0] = (-1)^{\ell} (2 \ell + 1) 
\pmatrix{
\ell & k & \ell \cr
   0 & 0 &    0 \cr
}
\sum_{q = -k}^{k} B^k_q ( k q | k a_0 \Gamma_0 \gamma_0)^*
      \label{connection D-B}
      \end{eqnarray} 
and $a_0$ is a branching multiplicity label to be used when $\Gamma_0$ appears several times in the decomposition of the $(k)$ IRC 
of $SO(3)$. (The index $\gamma_0$ in (\ref{symmetry-adapted form of Hcf}) and (\ref{connection D-B}) is not really necessary since 
$\Gamma_0$ is a one-dimensional IRC; it is mentioned only for aesthetic reasons.) Then, the matrix elements of ${\cal H}_{cf}$ in a 
$SU(2) \supset G^*$ symmetry-adapted weak-field basis are given by 
      \begin{eqnarray}
& & ( n \ell^N \alpha SLJ a \Gamma \gamma | {\cal H}_{cf} | n \ell^N \alpha' S' L' J' a' \Gamma' \gamma' ) = \delta_{S S'} 
\delta_{\Gamma  \Gamma'} \delta_{\gamma  \gamma'}
			\nonumber \\ 
& \times & (-1)^{S + L' + J} \sqrt{ (2J+1) (2J'+1) } \sum_{k a_0} D[k a_0] 
			\label{H cf} \\ 
& \times & ( n \ell^N \alpha S L \Vert U^{(k)} \Vert n \ell^N \alpha' S L' ) 
\left\lbrace\matrix{
L & k &L'\cr 
J'& S &J \cr
}\right\rbrace
f
\pmatrix{
J        & J'        & k            \cr
a \Gamma & a' \Gamma & a_0 \Gamma_0 \cr
}
      \nonumber
      \end{eqnarray}
where $\{ \cdots \}$ stands for a 6--$j$ Wigner symbol and $f$ is a coupling coefficient defined by 
      \begin{eqnarray}
f
\pmatrix{
J        & J'         & k            \cr
a \Gamma & a' \Gamma  & a_0 \Gamma_0 \cr
} & := &  
\sum_{M=-J}^{J} \sum_{M'=-J'}^{J'} \sum_{q=-k}^{k} 
( JM | J a \Gamma \gamma )^* ( kq | k a_0 \Gamma_0 \gamma_0 ) ( J'M' | J' a' \Gamma \gamma ) 
\nonumber \\
& \times & (-1)^{J-M} 
\pmatrix{
J   & k  & J' \cr
-M  & q  & M' \cr
}
      \label{f Gamma zero}
      \end{eqnarray}
This $f$ coefficient is independent of $\gamma$ \cite{Flato2,JMS68}. It 
is a particular case of the $f$ coefficient defined in \cite{JMS68} by
      \begin{eqnarray}
f
\pmatrix{
j_1    &  j_2    & k     \cr
\mu_1  &  \mu_2  & \mu   \cr
} &:=& 
\sum_{m_1=-j_1}^{j_1} \sum_{m_2=-j_2}^{j_2} \sum_{q=-k}^{k} 
( j_1m_1 | j_1 \mu_1 )^* ( kq | k \mu ) ( j_2m_2 | j_2 \mu_2 )
\nonumber \\ 
& \times & (-1)^{j_1-m_1} 
\pmatrix{
j_1   & k  & j_2 \cr
-m_1  & q  & m_2 \cr
} 
      \label{f general}
      \end{eqnarray}
where
      \begin{eqnarray}
\mu_1 := a_1 \Gamma_1 \gamma_1, \quad \mu_2 := a_2 \Gamma_2 \gamma_2, \quad \mu:= a \Gamma \gamma
      \label{crystalographic number}
      \end{eqnarray}
(see also \cite{kibler79}). As a conclusion, the calculation of the matrix of ${\cal H}$ in a symmetry-adapted 
weak-field basis (via (\ref{H Coulomb}), (\ref{H spin-orbit}) and (\ref{H cf})) is considerably 
simpler than in a strong-field basis. 

In the case of $d^N$ in $O$, the just described symmetry-adapted weak-field approach, based on 
(\ref{H Coulomb}), (\ref{H spin-orbit}) and (\ref{H cf}), leads to a weak-field model 
which is equivalent to the ordinary (or ionic) strong-field model with the parameters $A$, $B$, $C$, $\zeta_{nd}$ 
and $10Dq$. More generally for $\ell^N$ in $G$, the symmetry-adapted weak-field model and the ionic strong-field 
model are equivalent. However for $\ell^N$ in $G$, the symmetry-adapted weak-field model and the generalized (or 
covalent) strong-field model are not equivalent. Thus, it is desirable to develop a generalized symmetry-adapted 
weak-field model equivalent to the generalized strong-field model. This will be done in the next section. 

\subsection{Generalized weak-field model}

To generalize the symmetry-adapted weak-field model, we keep the symmetry-adapted weak-field basis 
(\ref{weak-field vectors}) intact in order to take advantage of its simplicity. The sole modification 
to be done consists in replacing the Hamiltonian ${\cal H}$ by an effective Hamiltonian $H_{\rm eff}$. The 
Hamiltonian $H_{\rm eff}$ for $\ell^N$ in $G$ should reduce to ${\cal H}$ for some special values of its 
parameters, should be an Hermitian operator invariant under the $G$ group and the time-reversal operator, 
and should contain one- and two-body spin and orbit interactions. For the sake of easy 
calculations, $H_{\rm eff}$ should involve a coupling scheme which is reminiscent of the $\{ SLJ \}$ coupling 
scheme of the state vectors (\ref{weak-field vectors}). Therefore, we take $H_{\rm eff}$ in the form
      \begin{eqnarray}
H_{\rm eff} & := & \sum_{i , j} \sum_{{ \rm all \, } k} \sum_{a_0} D[(k_1k_2)k_S (k_3k_4)k_L k a_0] 
\nonumber \\
& \times & \{ \{ {\bf u}^{(k_1)}(i) \otimes {\bf u}^{(k_2)}(j) \}^{(k_S)} \otimes 
              \{ {\bf u}^{(k_3)}(i) \otimes {\bf u}^{(k_4)}(j) \}^{(k_L)} \}^{(k)}_{a_0 \Gamma_0 \gamma_0}
      \label{the Hamiltonian H}
      \end{eqnarray}
where the ${\bf u}$'s are one-electron Racah unit tensor operators with 
$\{ {\bf u}^{(k_1)}(i) \otimes {\bf u}^{(k_2)}(j) \}^{(k_S)}$ acting on the spin part and 
$\{ {\bf u}^{(k_3)}(i) \otimes {\bf u}^{(k_4)}(j) \}^{(k_L)}$ on the orbital part of the state vectors  
(\ref{weak-field vectors}). The sums over $i$ and $j$ in (\ref{the Hamiltonian H}) are extended over the $N$ electrons 
and the sums over the $k$'s and $a_0$ are limited, like in (\ref{symmetry-adapted form of Hcf}), by hermiticity and 
symmetry properties of $H_{\rm eff}$ (invariance under the $G$ group and the time-reversal operator) and by the selection rules on the 
matrix elements of $H_{\rm eff}$ in the basis (\ref{weak-field vectors}). Furthermore, the parameters $D[(k_1k_2)k_S (k_3k_4)k_L k a_0]$ 
comprise the Coulomb interelectronic parameters, the spin-orbit parameters and the crystal-field parameters of the ordinary 
weak-field model plus some additional parameters to be described below. The most important (as far as a comparison with the generalized strong-field model is in order) parameters in $H_{\rm eff}$ can be classified in the following way.
\begin{enumerate}
	\item The $D[(00)0 (kk)0 0]$ parameters correspond to the ordinary or isotropic Coulomb interaction between the $N$ electrons.
	\item The $D[(00)0 (k_3k_4)k_L k_L a_0]$ parameters with $k_L \not= 0$ correspond to anisotropic Coulomb interactions between the $N$ electrons 
	or ligand-field correlated Coulomb interactions.
	\item The $D[(ss)1 (\ell \ell)1 0]$ parameter corresponds to the ordinary or isotropic spin-orbit interaction for the $N$ electrons ($s = 1/2$).
	\item The $D[(ss)1 (\ell \ell)k_L k a_0 ]$ parameters with $k_L \not= 1$ correspond to anisotropic spin-orbit interactions for the $N$ electrons 
	or ligand-field correlated spin-orbit interactions.
	\item The $D[(ss)0 (\ell \ell)k_L k_L a_0]$ parameters correspond to the ligand-field interaction.
\end{enumerate}

The building of the energy matrix of $H_{\rm eff}$ in the basis (\ref{weak-field vectors}) is very 
simple. Indeed, we have the following matrix elements
      \begin{eqnarray}
& & ( n \ell^N \alpha SLJ a \Gamma \gamma | H_{\rm eff} | n \ell^N \alpha' S' L' J' a' \Gamma' \gamma' ) = 
\delta_{\Gamma  \Gamma'} \delta_{\gamma  \gamma'}
\nonumber \\
& \times & \sum_{{ \rm all \, } k} \sum_{a_0} D[(k_1k_2)k_S (k_3k_4)k_L k a_0] 
f \pmatrix{
J        & J'        & k            \cr
a \Gamma & a' \Gamma & a_0 \Gamma_0 \cr
} \sum_{i , j} 
      \label{matrix of H} \\
& \times & 
(n\ell^N \alpha SLJ \Vert \{ \{ {\bf u}^{(k_1)}(i)\otimes {\bf u}^{(k_2)}(j) \}^{(k_S)} \otimes\{ 
                                {\bf u}^{(k_3)}(i)\otimes {\bf u}^{(k_4)}(j) \}^{(k_L)} \}^{(k)} \Vert n\ell^N\alpha' S'L'J') 
\nonumber 
      \end{eqnarray}
where the reduced matrix element $( \Vert \cdots \Vert )$ 
can be calculated from the Racah's standard methods. 

The symmetry-adapted weak-field approach based on (\ref{the Hamiltonian H}) and (\ref{matrix of H}) 
leads to a model that turns out to be equivalent to the generalized strong-field model. However, the 
generalized symmetry-adapted weak-field model contains more parameters than the generalized strong-field model
(e.g., the Hamiltonian given by (\ref{the Hamiltonian H}) contains spin-spin and orbit-orbit interaction parameters 
that do not occur in the generalized strong-field model). The $D[(k_1k_2)k_S (k_3k_4)k_L k a_0]$ 
parameters can be considered as phenomenological global parameters to be fitted on experimental data. All 
or part of these parameters can be interpreted and calculated in the framework of {\em ab initio} 
microscopic models as for instance the angular overlap model \cite{AOMJorg, AOMKibl}, the superposition model 
\cite{Newman} and the MO-LCAO model \cite{Bird, Pueyo, Daul1, Daul2}. (See the appendix for the connection 
between the isotropic Coulomb interaction parameters and the Slater-Condon-Shortley parameters.) Of course, the  
generalized symmetry-adapted weak-field model gives back the ordinary symmetry-adapted weak-field model as a 
particular case when some parameters vanish. 

By way of illustration, let us consider the case of $d^N$ in $O$. The corresponding Hamiltonian $H_{\rm eff}$ 
can be restricted to an operator containing 14 parameters, namely,  
\begin{itemize}
	\item 10 Coulomb parameters: \\
	$D[(00)0 (00)0 0]$,
	$D[(00)0 (22)0 0]$, 
	$D[(00)0 (44)0 0]$, \\
	$D[(00)0 (04)4 4]$,
	$D[(00)0 (22)4 4]$, 
	$D[(00)0 (24)4 4]$, \\
	$D[(00)0 (44)4 4]$,
	$D[(00)0 (24)6 6]$,
	$D[(00)0 (44)6 6]$,
	$D[(00)0 (44)8 8]$;
	\item 2 spin-orbit parameters: \\
	$D[(ss)1 (22)1 0]$,
	$D[(ss)1 (22)3 4]$;
	\item 2 ligand-field parameters: \\
	$D[(ss)0 (22)0 0]$,
	$D[(ss)0 (22)4 4]$;
\end{itemize}
It can be shown that the generalized symmetry-adapted weak-field model with these 14 parameters 
is equivalent to the generalized strong-field model for $d^N$ in $O$ \cite{IJQC5}. Such an 
equivalence was also worked out for the case of $f^N$ in $O$. In this case, the generalized 
symmetry-adapted weak-field model can be restricted to involve the following 33 parameters  
\begin{itemize}
	\item 26 Coulomb parameters: \\
	$D[(00)0 (00)0 0]$,
	$D[(00)0 (22)0 0]$, 
	$D[(00)0 (44)0 0]$,
	$D[(00)0 (66)0 0]$, \\
	$D[(00)0 (04)4 4]$, 
	$D[(00)0 (22)4 4]$, 
	$D[(00)0 (24)4 4]$,
	$D[(00)0 (26)4 4]$, \\
	$D[(00)0 (44)4 4]$,
	$D[(00)0 (46)4 4]$, 
	$D[(00)0 (66)4 4]$,
	$D[(00)0 (06)6 6]$, \\
	$D[(00)0 (24)6 6]$,
	$D[(00)0 (26)6 6]$, 
	$D[(00)0 (44)6 6]$, 
	$D[(00)0 (46)6 6]$, \\
	$D[(00)0 (66)6 6]$,
	$D[(00)0 (26)8 8]$, 
	$D[(00)0 (44)8 8]$,
	$D[(00)0 (46)8 8]$, \\
	$D[(00)0 (66)8 8]$,
	$D[(00)0 (46)9 9]$, 
	$D[(00)0 (46)10, 10]$,
	$D[(00)0 (66)10, 10]$, \\
	$D[(00)0 (66)12, 12a]$, 
	$D[(00)0 (66)12, 12b]$;
	\item 4 spin-orbit parameters: \\
	$D[(ss)1 (33)1 0]$,
	$D[(ss)1 (33)3 4]$, 
	$D[(ss)1 (33)5 4]$,
	$D[(ss)1 (33)5 6]$;
	\item 3 ligand-field parameters: \\
	$D[(ss)0 (33)0 0]$,
	$D[(ss)0 (33)4 4]$, 
	$D[(ss)0 (33)6 6]$.
\end{itemize}
The generalized symmetry-adapted weak-field model with these 33 parameters 
is equivalent to the generalized strong-field model for $f^N$ in $O$ \cite{IJQC5}.  

\subsection{Transition intensities}

In addition to be useful for the calculation of energy levels of a partly-filled shell ion in a given surrounding, 
the Racah's methods proved to be of considerable importance for the calculation of transitions between levels. We 
shall not develop these facets of crystal- and ligand-field theory here. It is enough to mention the pioneer works 
by Judd \cite{Judd1962} and Ofelt \cite{Ofelt1962} for one-photon electric dipolar transitions between split levels 
of the same parity (see also \cite{Wybournebook}). Let us also mention that the symmetry considerations developed 
by Bader and Gold \cite{BaderGold1968} for two-photon electric dipolar transitions between states of opposite 
parities were reformulated in the symmetry-adapted weak-field model \cite{CCA6, KibDao93, DaoKib95}. Finally, let 
us mention that irreducible tensor methods for finite groups were used for calculating the intensities of 
photoelectron spectra of partly-filled shell ion systems \cite{Cox, Grenet1, Grenet2, Grenet3, Grenet4}.





\section{Wigner-Racah algebra for a finite or compact group}
 
An important task in spectroscopy is to calculate matrix elements in order 
to determine energy spectra and transition intensities. In the case of 
many-fermionic systems, this can be done either in the Slater-Condon-Shortley 
approach (with determinantal states) or in the Dirac-Wigner-Racah approach 
(with states characterized by quantum numbers). In the Dirac-Wigner-Racah approach, 
one way to incorporate symmetry considerations connected to a chain of groups (involving symmetry groups and 
classification groups) is to use the `Wigner-Racah calculus' associated with the chain under consideration. The 
`Wigner-Racah calculus' or `Wigner-Racah algebra' associated with a group $G$ (or a chain of groups $G_{a} \supset G_{\Gamma}$) 
is generally understood as the set of algebraic manipulations concerning the 
coupling and recoupling coefficients for the group $G$ (or the head group $G_{a}$). This `algebra' may be also understood as a true 
algebra in the mathematical sense: It is the (in)finite-dimensional Lie algebra spanned by the irreducible 
unit tensor operators or Wigner operators of $G$ (or $G_{a}$) \cite{IASP, Springer, KibGreJMP, BLangular, BLWRa}. We 
shall mainly focus here on the very basic aspects of the `algebra' of the coupling and recoupling coefficients 
of a finite or compact group $G$. The Wigner-Racah calculus was originally developed for simply-reducible 
(i.e., ambivalent plus multiplicity-free) groups \cite{wigner41, wigner65, wigner68}. (Let us recall that a group 
$G$ is said to be ambivalent if each element of $G$ and its inverse belong to a same conjugation class. It is said 
to be multiplicity-free if the Kronecker product of two arbitrary irreducible representations of $G$ contains at 
most once each irreducible representation of $G$.) The bases of the Wigner-Racah algebra of the rotation group, 
a simply-reducible group, were introduced at the beginning of the forties by Wigner \cite{wigner65} and Racah 
\cite{RII,RIII}. In the sixties and seventies, the idea of a Wigner-Racah algebra was extended to an arbitrary 
finite or compact group \cite{WTSharp, Derome1, Derome2} (see the review in \cite{Butler}) and started to be 
applied to some groups or chains of groups of interest in crystal- and ligand-field theory 
\cite{stk70,griffith62,chinois2,Smirnov5,kibler79}. Regarding molecular and solid-state physics, let us also 
mention that Koster {\it et al.} published the first complete set of tables of coupling coefficients for the 
thirty-two (single and double) crystallographic point groups \cite{Koster}. Most of the developments concerning 
chains of groups were strongly influenced by a lemma due to Racah derived in \cite{RIV} for an arbitrary chain 
involving finite and/or compact groups. 

We present in what follows the basic ingredients for the Wigner-Racah algebra of a finite or compact group 
in a terminology easily adaptable to nuclear, atomic, molecular, and condensed matter physics as well as in 
quantum chemistry.


\subsection{Preliminaries}
Let us consider an arbitrary finite or compact continuous group $G$ having 
the IRCs $a$, $b$, $\cdots$. The identity IRC, often noted $A$ or $A_1$ or 
$\Gamma_1$ in molecular physics, is denoted by 0 in this section (it is 
noted $\Gamma_0$ in Section 2). To each IRC $a$, we associate a unitary matrix 
representation $D^a$. Let [a] be the dimension of $D^a$. The 
$\alpha$-$\alpha'$ matrix element of the representative $D^a(R)$ for 
the element $R$ in $G$ is written $D^a(R)_{\alpha\alpha'}$. (For $a=0$, 
we have $\alpha = \alpha' = 0$.) The sum $\chi^a(R) = \sum_{\alpha} 
D^a(R)_{\alpha\alpha}$ stands for the character of $R$ in 
$D^a$. The $D^a(R)_{\alpha\alpha'}$ and $\chi^a(R)$ satisfy 
orthogonality relations (e.g., the so-called great orthogonality theorem for $D^a(R)_{\alpha\alpha'}$) 
that are very familiar to the physicist and the chemist. We use $\left\vert G \right\vert$ 
to denote the order of $G$ when $G$ is a finite group or the volume $\int_G dR$ of $G$ when $G$ is 
a compact continuous group. Furthermore, the notation $\int_G \ldots dR$, which applies when $G$ is  
a compact continuous group, should be understood as $\sum_{R \in G} \ldots$ when $G$ is a 
finite group. 

\subsection{Clebsch-Gordan coefficients}

The direct product $a \otimes b$ of two IRCs $a$ and $b$ of $G$ can be in 
general decomposed into a direct sum of IRCs of $G$. This leads to the 
Clebsch-Gordan series
\begin{eqnarray}
a \otimes b = \bigoplus_c \sigma (c | a \otimes b) c
\label{eq1}
\end{eqnarray}
where $\sigma (c | a \otimes b)$ denotes the number of times the $c$ IRC 
occurs in $a \otimes b$. The integers $\sigma (c | a \otimes b)$ may be 
determined through the character formula
\begin{eqnarray}
       \sigma (c | a \otimes b) = 
\left\vert G \right\vert^{-1}{\int_G} \chi^c(R)^* \chi^a(R) \chi^b(R) dR
\label{eq2}
\end{eqnarray}
In terms of matrix representations, (\ref{eq1}) reads
\begin{eqnarray}
D^a \otimes D^b \sim 
                  \bigoplus_c \sigma (c | a \otimes b) D^c
\label{eq3}
\end{eqnarray}
Therefore, there exists a unitary matrix ${U^{ab}}$ such that
\begin{eqnarray}
({U^{ab}})^\dagger D^a(R) \otimes D^b(R) {U^{ab}} = \bigoplus_c 
\sigma (c | a \otimes b) D^c(R)
\label{eq4}
\end{eqnarray}
or equivalently
\begin{eqnarray}
D^a(R) \otimes D^b(R) = \bigoplus_c \sigma (c | a \otimes b) 
{U^{ab}} D^c(R) ({U^{ab}})^\dagger 
\label{eq5}
\end{eqnarray}
for any $R$ in $G$. It is a simple exercise in linear algebra to transcribe 
(\ref{eq4}) and (\ref{eq5}) in matrix elements. We thus have
\begin{eqnarray}
\sum_{\alpha \beta \alpha' \beta'} 
\left( {U^{ab}} \right)_{\alpha  \beta ,\rho  c  \gamma}^*
D^a(R)_{\alpha \alpha'}
D^b(R)_{\beta   \beta'}     
\left( {U^{ab}} \right)_{\alpha' \beta',\rho' c' \gamma'}
= \Delta (c \vert a \otimes b) \delta _{\rho\rho'} \delta _{cc'}
D^c(R)_{\gamma \gamma'}
\label{new6}
\end{eqnarray}
and
\begin{eqnarray}
D^a(R)_{\alpha \alpha'} D^b(R)_{\beta   \beta'}
= \sum_{\rho c \gamma \gamma'} 
\left( {U^{ab}} \right)_{\alpha  \beta ,\rho c \gamma}
D^c(R)_{\gamma \gamma'} 
\left( {U^{ab}} \right)_{\alpha' \beta',\rho c \gamma'}^*
\label{new7}
\end{eqnarray}
for any $R$ in $G$. Each row index of ${U^{ab}}$ consists of two labels 
($\alpha$ and $\beta$) according to the rules of the direct product 
of two matrices. Similarly, two labels ($c$ and $\gamma$) are required
for characterizing each column index of ${U^{ab}}$. However, when $c$ 
appears several times in $a \otimes b$, a third label (the multiplicity 
label $\rho$) is necessary besides $c$ and $\gamma$. Hence, the summation 
over $\rho$ in (\ref{new7}) ranges from 1 to $\sigma (c \vert a \otimes b)$. Finally 
in (\ref{new6}), $\Delta (c \vert a \otimes b) = 0$ or 1 according to whether as 
$c$ is contained or not in $a \otimes b$. (Note that $\Delta (c \vert a \otimes b)$ 
is the analog of $\Delta (S, L, J)$ used in Section 2.)

Following the tradition in quantum mechanics, we put 
\begin{eqnarray}
 (a b \alpha \beta \vert \rho c \gamma) 
                           := \left( {U^{ab}} \right)_{\alpha \beta,\rho c \gamma}
\label{eq8}
\end{eqnarray}
so that (\ref{new6}) and (\ref{new7}) can be rewritten as 
\begin{eqnarray}
\sum_{\alpha \beta \alpha' \beta'} (a b \alpha \beta \vert \rho c \gamma )^*
D^a(R)_{\alpha \alpha'}
D^b(R)_{\beta   \beta'}     (a b \alpha'\beta'\vert \rho'c'\gamma')
= \Delta (c \vert a \otimes b) \delta _{\rho\rho'} \delta _{cc'}
D^c(R)_{\gamma \gamma'}
\label{eq6}
\end{eqnarray}
and
\begin{eqnarray}
D^a(R)_{\alpha \alpha'} D^b(R)_{\beta   \beta'}
= \sum_{\rho c \gamma \gamma'} (a b \alpha \beta \vert \rho c \gamma )
D^c(R)_{\gamma \gamma'} (a b \alpha'\beta'\vert \rho c \gamma')^* 
\label{eq7}
\end{eqnarray}
The matrix elements $(a b \alpha \beta \vert \rho c \gamma)$ are termed 
Clebsch-Gordan coefficients (CGCs) or vector coupling coefficients. The 
present introduction clearly emphasizes that the CGCs of a group $G$ are 
nothing but the elements of a unitary matrix which reduces the direct product 
of two irreducible matrix representations of $G$. As a consequence, the CGCs 
satisfy two orthonormality relations associated with the unitary property of 
${U^{ab}}$:
\begin{eqnarray}
\sum_{\alpha \beta} (a b \alpha \beta \vert \rho c \gamma )^* 
                    (a b \alpha \beta \vert \rho'c'\gamma') = 
\Delta(c \vert a \otimes b) 
\delta_{\rho \rho'} \delta_{c c'} \delta_{\gamma \gamma'}
\label{eq9}
\end{eqnarray}
and
\begin{eqnarray}
\sum_{\rho c \gamma} (a b \alpha \beta \vert \rho c \gamma)
                     (a b \alpha'\beta'\vert \rho c \gamma)^* = 
\delta_{\alpha \alpha'} \delta_{\beta \beta'}
\label{eq10}
\end{eqnarray}
Note that (\ref{eq9}) and (\ref{eq10}) are conveniently recovered by specializing 
$R$ to the unit element $E$ of $G$ in (\ref{eq6}) and (\ref{eq7}), respectively. As 
an evident selection rule on the CCGs, it is clear that in order to have 
$(a b \alpha \beta \vert \rho c \gamma) \not=0$ it is necessary (but not sufficient) 
that $c$ be contained in $a \otimes b$.

Equations (\ref{eq6}) and (\ref{eq7}) show that the CGCs are basis-dependent 
coefficients. In this regard, it is important to realize that (\ref{eq6}) and 
(\ref{eq7}) are not sufficient to define unambiguously the CGCs of the $G$ group 
once its irreducible representation matrices are known. As a matter of fact, the 
relation
		\begin{eqnarray}
(a b \alpha \beta \vert r    c \gamma) := \sum_{\rho} 
(a b \alpha \beta \vert \rho c \gamma) M(a b,c)_{\rho r}
\label{eq11}
		\end{eqnarray}
where $M(ab,c)$ is an arbitrary unitary matrix of dimension 
$\sigma (c \vert a \otimes b) \times \sigma (c \vert a \otimes b)$, 
defines a new set of CGCs since
(\ref{eq6}) and (\ref{eq7}) are satisfied by making replacements of type 
$\rho \to r$. The CGCs associated with a definite choice for the irreducible 
representation matrices of $G$ are thus defined up to a unitary transformation, 
a fact that may be exploited to generate special symmetry properties of the CGCs.

Various relations involving elements of irreducible representation matrices and
CGCs can be derived from (\ref{eq6}) and (\ref{eq7}) by using the unitarity property 
both for the representation matrices and the Clebsch-Gordan matrices. For instance,
we obtain
\begin{eqnarray}
\sum_{\alpha' \beta'}D^a(R)_{\alpha \alpha'}
                     D^b(R)_{\beta  \beta '}
(a b \alpha' \beta' \vert \rho c \gamma') = \sum_{\gamma} 
(a b \alpha  \beta  \vert \rho c \gamma ) D^c(R)_{\gamma \gamma'}
\label{eq12} 
\end{eqnarray}
\begin{eqnarray}
\sum_{\alpha'}D^a(R)_{\alpha \alpha'}
(a b \alpha' \beta' \vert \rho c \gamma') = \sum_{\beta \gamma} 
(a b \alpha  \beta  \vert \rho c \gamma ) 
D^b(R)_{\beta  \beta '}^*
D^c(R)_{\gamma \gamma'}
\label{entre12et13}
\end{eqnarray}
\begin{eqnarray}
(a b \alpha' \beta' \vert \rho c \gamma') = \sum_{\alpha \beta \gamma}
(a b \alpha  \beta  \vert \rho c \gamma )
D^a(R)_{\alpha \alpha'}^*
D^b(R)_{\beta  \beta '}^*
D^c(R)_{\gamma \gamma'}
\label{eq13}
\end{eqnarray}
for any $R$ in $G$. In the situation where the elements of the irreducible
representation matrices of $G$ are known, Eqs.~(\ref{eq12}), (\ref{entre12et13}) 
and (\ref{eq13}) provide us with linear equations useful for checking the 
numerical values of the CGCs of $G$. 

The combination of (\ref{eq7}) with the great orthogonality theorem for $G$ 
yields the relation
\begin{eqnarray}
\vert G \vert^{-1} \int_G 
D^a(R)_{\alpha \alpha'}
D^b(R)_{\beta  \beta '}
D^c(R)_{\gamma \gamma'}^* dR
= [c]^{-1} \sum_{\rho} 
(a b \alpha  \beta  \vert \rho c \gamma) 
(a b \alpha' \beta' \vert \rho c \gamma')^*
\label{eq14}
\end{eqnarray}
which is useful for the calculation of the CGCs of $G$ in terms of the
elements of the irreducible representation matrices of $G$. Note that when
$a \otimes b$ is multiplicity-free (i.e., when there is no summation on 
$\rho$ in (\ref{eq14})), Eq.~(\ref{eq14}) allows us to determine 
$(a b \alpha \beta \vert c \gamma)$ for all $\alpha$, $\beta$ and $\gamma$ 
up to arbitrary phase factors~; more precisely, we then have
\begin{eqnarray}
(a b \alpha \beta | c \gamma) = {\rm e}^{{\rm i} h(ab,c)} \left( \frac{ [c] }{ |G| } \right)^{1/2}
\frac{
\int_G   D^a(R)_{\alpha \alpha'} 
         D^b(R)_{\beta  \beta '}
         D^c(R)_{\gamma \gamma'}^* dR}
{\lbrace  
\int_G   D^a(R)_{\alpha' \alpha'}
         D^b(R)_{\beta ' \beta '}
         D^c(R)_{\gamma' \gamma'}^* dR \rbrace^{ 1/2 } }
\label{eq15}
\end{eqnarray}
where $h(ab,c) \in \mathbb{R}$. 

It appears from (\ref{eq12})-(\ref{eq15}) that $c$ does not generally play the same role 
as $a$ and $b$ in $(a b \alpha \beta \vert \rho c \gamma)$. Indeed, (\ref{eq13}) shows that 
the CGCs $(a b \alpha \beta | \rho c \gamma)$ are the components of a third rank tensor, twice 
contravariant and once covariant. Therefore, 
               $(a b \alpha \beta \vert \rho c \gamma)$ does not generally 
exhibit simple symmetry properties under permutations of $a$, $b$ and $c$. It 
will be shown in the following how the CGCs may be symmetrized thanks to a 
2--$a \alpha$ symbol.

\subsection{The 2--$a \alpha$ symbol}

Let us define  the 2--$a \alpha$ symbol through
		\begin{eqnarray}
\pmatrix{
a&b\cr
\alpha&\beta\cr
} := [a]^{1/2} (b a \beta \alpha \vert 00)
		\label{eq16}
		\end{eqnarray}
The 2--$a \alpha$ symbol makes it possible to pass from a given irreducible
matrix representation to its complex conjugate. This is reflected by the two
relations
\begin{eqnarray}
\sum_{\alpha \alpha'} \pmatrix{
a&b\cr
\alpha&\beta\cr
}^*
D^a(R)_{\alpha\alpha'} \pmatrix{
a&b'\cr
\alpha'&\beta'\cr
}
= \Delta (0 \vert a \otimes b) \delta_{b b'} D^b(R)_{\beta \beta '}^*
\label{eq17}
\end{eqnarray}
and
    \begin{eqnarray}
\sum_{\beta \beta'} \pmatrix{
a&b\cr
\alpha&\beta\cr
} 
D^b(R)_{\beta\beta'}^* \pmatrix{
a'&b\cr
\alpha'&\beta'\cr
}^*
= \Delta (0 \vert a \otimes b) \delta_{a a'} D^a(R)_{\alpha\alpha'}
    \label{eq18}
    \end{eqnarray}
that hold for any $R$ in $G$. The proof of (\ref{eq17}) and (\ref{eq18}) is long~; it 
starts with the introduction of (\ref{eq16}) into the left-hand sides of (\ref{eq17})
and (\ref{eq18}) and requires repeated use of relations involving the irreducible matrix 
representations and CGCs as well as the great orthogonality theorem of $G$. By taking 
$R = E$ in (\ref{eq17}) and (\ref{eq18}), we get the 
useful relations
\begin{eqnarray}
\sum_{\alpha}\pmatrix{
a&b\cr
\alpha&\beta\cr
}^*
\pmatrix{
a&b'\cr
\alpha&\beta'\cr
}
= \Delta (0\vert a\otimes b) \delta_{bb'} \delta_{\beta\beta'}
\label{eq19}
\end{eqnarray}
and
\begin{eqnarray}
\sum_{\beta}\pmatrix{
a&b\cr
\alpha&\beta\cr
}
\pmatrix{
a'&b\cr
\alpha'&\beta\cr
}^*
 = \Delta (0\vert a\otimes b) \delta_{aa'} \delta_{\alpha\alpha'}
\label{eq20}
\end{eqnarray}
which give back (\ref{eq9}) as particular case. 

The  2--$a \alpha$ symbol turns out to be of relevance for handling phase
problems. In this regard, both (\ref{eq17}) and (\ref{eq18}) lead to
\begin{eqnarray}
\delta_{ab} \sum_{ \alpha \beta } \pmatrix{
a & b \cr
\alpha & \beta  \cr
}^*
\pmatrix{
b & a \cr
\beta  & \alpha \cr
}
= \Delta (0 \vert a \otimes b) [a] c_a
\label{eq21}
\end{eqnarray}
where the Frobenius-Schur coefficient
    \begin{eqnarray}
c_a := \vert G \vert^{-1} \int_G \chi^a (R^2) dR
    \label{eq22}
    \end{eqnarray}
is 1, $-1$, or 0 according to as $D^a$ is orthogonal, 
symplectic, or complex (i.e., integer, half-integer or complex in Wigner's terminology). Note that 
\begin{eqnarray}
c_a \pmatrix{
b & a \cr
\beta & \alpha \cr
}
= \delta_{ab} \pmatrix{
a & b\cr
\alpha & \beta \cr
}
\label{eq23}
\end{eqnarray}
satisfies (\ref{eq21}). Equation (\ref{eq23}) reflects the symmetry of the matrix which 
enables to pass from the matrix $D^a$ to its complex conjugate $( D^a )^*$ 
(cf., the Frobenius-Schur theorem). Thus, the 2--$a \alpha$ symbol plays the role of a metric tensor 
that transforms $D^a$ into $( D^a )^*$. It generalizes the Herring-Wigner metric tensor 
introduced for the $SU(2)$ group (see \cite{wigner65}).  

\subsection{The $($3--$a \alpha)_{\rho}$ symbol}

We now define the $($3--$a \alpha)_\rho$ symbol via
		\begin{eqnarray}
\pmatrix{
a&b&c\cr
\alpha&\beta&\gamma\cr
}_\rho
:= \sum_{\rho' c' \gamma'} [c']^{-1/2} M(ba,c')_{\rho' \rho} \pmatrix{
c&c'\cr
\gamma&\gamma'\cr
}
(b a \beta \alpha \vert \rho' c' \gamma')
		\label{eq24}
		\end{eqnarray}
where $M(ba,c')$ is an arbitrary unitary matrix. Conversely, each CGC can 
be developed in terms of $($3--$a \alpha)_\rho$ symbols since the inversion 
of (\ref{eq24}) gives 
\begin{eqnarray}
(a b \alpha \beta \vert \rho c \gamma) = 
[c]^{1/2} \sum_{\rho' c' \gamma'}
M(ab,c)_{\rho\rho'}^*\pmatrix{
c'&c\cr
\gamma'&\gamma\cr
}^*
\pmatrix{
b&a&c'\cr
\beta&\alpha&\gamma'\cr
}_{\rho'}
\label{eq25}
\end{eqnarray}
after utilization of the unitarity property of the 2--$a \alpha$ symbol 
and of the matrix $M(ba,c')$. 

All the relations involving CGCs may be transcribed in terms of
$($3--$a\alpha)_\rho$ symbols. For example, the orthonormality relations 
(\ref{eq9}) and (\ref{eq10}) are easily amenable to the form
 \begin{eqnarray}
\sum_{\rho c \gamma} [c] \pmatrix{
a&b&c\cr
\alpha&\beta&\gamma\cr
}_\rho
\pmatrix{a&b&c\cr
\alpha'&\beta'&\gamma\cr
}_\rho^*
= \delta_{\alpha \alpha'} \delta_{\beta \beta'}
\label{eq29}
\end{eqnarray}
and 
\begin{eqnarray}
\sum_{\alpha \beta} \pmatrix{
a&b&c\cr
\alpha&\beta&\gamma\cr
}_\rho^*
\pmatrix{
a&b&c'\cr
\alpha&\beta&\gamma'\cr
}_{\rho'}
= \Delta (0 \vert a \otimes b \otimes c) \delta_{\rho \rho'} \delta_{c c'}
\delta_{\gamma \gamma'} [c]^{-1}
\label{eq28}
\end{eqnarray} 
Along the same line, the introduction of (\ref{eq25}) into (\ref{eq7}) yields 
\begin{eqnarray}
D^a(R)_{\alpha\alpha'}D^b(R)_{\beta\beta'} = 
\sum_{\rho c \gamma \gamma'} [c] 
\pmatrix{
a&b&c\cr
\alpha&\beta&\gamma\cr
}_\rho
D^c(R)_{\gamma\gamma'}^*\pmatrix{
a&b&c\cr
\alpha'&\beta'&\gamma'\cr
}_\rho^*
\label{eq27}
\end{eqnarray}
which in turn leads to 
\begin{eqnarray}
 \sum_{\alpha \beta \alpha' \beta'} \pmatrix{
a&b&c\cr
\alpha&\beta&\gamma\cr
}^*_\rho
D^a(R)_{\alpha \alpha'}
D^b(R)_{\beta  \beta' }
\pmatrix{
a&b&c'\cr
\alpha'&\beta'&\gamma'\cr
}_{\rho'} 
 \cr \cr
 = \Delta (0 \vert a \otimes b \otimes c) \delta_{\rho \rho'} \delta_{c c'} 
[c]^{-1} D^c(R)_{\gamma\gamma'}^* 
\label{eq26}
\end{eqnarray}
owing to the orthogonality relation (\ref{eq28}). Equations (\ref{eq27}) and (\ref{eq26}) 
hold for any element $R$ in $G$. As a check, note that for $R = E$, they can be specialized 
to (\ref{eq29}) and (\ref{eq28}). 

Relation (\ref{eq27}) and its dual relation (\ref{eq26}) show that $D^a$, $D^b$ 
and $D^c$ present the same variance. This may be precised by 
\begin{eqnarray}
\pmatrix{
a       & b      & c       \cr
\alpha' & \beta' & \gamma' \cr
}_\rho 
= \sum_{\alpha \beta \gamma}
\pmatrix{
a      & b     & c      \cr
\alpha & \beta & \gamma \cr
}_\rho
D^a(R)_{\alpha \alpha'}^*
D^b(R)_{\beta  \beta '}^*
D^c(R)_{\gamma \gamma'}^*
\label{variance}
\end{eqnarray}
which shows that the behavior of the $($3--$a \alpha)_\rho$ 
symbol under permutations of $a$, $b$ and $c$ should be easier to describe than the one of the 
CGC $(a b \alpha \beta \vert \rho c \gamma)$. This is reflected by the following relation (to 
be compared to (\ref{eq14})) 
\begin{eqnarray}
\vert G\vert^{-1}\int_G 
D^a(R)_{\alpha \alpha'}
D^b(R)_{\beta  \beta '}
D^c(R)_{\gamma \gamma'} dR
= \sum_{\rho} \pmatrix{
a&b&c\cr
\alpha&\beta&\gamma\cr
}_\rho
\pmatrix{
a&b&c\cr
\alpha'&\beta'&\gamma'\cr
}_\rho^* 
\label{eq30}
\end{eqnarray}
which may be proved directly by combining (\ref{eq27}) with the great orthogonality
theorem for the $G$ group. When the triple direct product 
$a \otimes b \otimes c$ contains the identity IRC of $G$ only once (i.e., when there 
is no label $\rho$ and no summation in (\ref{eq30})), Eq.~(\ref{eq30}) shows 
that the square modulus of the 3--$a \alpha$ symbol is invariant
under permutation of its columns. In this case, we may take advantage of the
arbitrariness of the matrix $M$ in (\ref{eq11}) or (\ref{eq24}) to produce 
convenient symmetry properties of the 3--$a \alpha$ symbol 
under permutations of its columns. By way of illustration, let us mention the
following result \cite{wigner65}: For $G$ simply reducible, it is possible to
arrange that the numerical value of the 3--$a \alpha$ symbol 
be multiplied by the phase factor $(-1)^{a+b+c}$, with $(-1)^{2 x } = c_x$, 
under an odd permutation of its columns~; consequently, the numerical value of
the 3--$a \alpha$ symbol remains unchanged under an even permutation of its 
columns (since $c_a c_b c_c = 1$). 

To close this subsection, we note that the $($3--$a \alpha)_\rho$ symbol constitutes 
a generalization to the case of an arbitrary finite or compact group of the 3--$jm$ 
symbol introduced by Wigner for simply reducible groups (in particular for the 
rotation group) \cite{wigner65} and of the 
$\overline{V}$ symbol introduced by Fano and Racah for the $SU(2)$ group 
\cite{FanoRacah} (the $\overline{V}$ symbol is a symmetrized version of the $V$ 
symbol defined by Racah \cite{RII}). 

\subsection{Recoupling coefficients}

We now define two new coefficients:  
		\begin{eqnarray}
(a (bc) \rho_{bc} c_{bc} \rho' d' \delta' \vert 
 (ab) \rho_{ab} c_{ab} c \rho  d  \delta) &:=&
\sum_{\alpha \beta \gamma                        } 
\sum_{                    \gamma_{ab} \gamma_{bc}} 
(a b \alpha \beta \vert \rho_{ab} c_{ab} \gamma_{ab})
(c_{ab} c \gamma_{ab} \gamma \vert \rho d \delta) 
\nonumber \\
&\times& (b c \beta \gamma \vert \rho_{bc} c_{bc} \gamma_{bc})^* 
       (a c_{bc} \alpha \gamma_{bc} \vert \rho' d' \delta')^*
		\label{eq31}
		\end{eqnarray}
and 
		\begin{eqnarray}
& & ((ac) \rho_{ac} c_{ac} (bd) \rho_{bd} c_{bd} \rho'e'\varepsilon'\vert
 (ab) \rho_{ab} c_{ab} (cd) \rho_{cd} c_{cd} \rho e \varepsilon)
		\nonumber \\
&:=& \sum_{\alpha \beta \gamma \delta}
    \sum_{\gamma_{ab} \gamma_{cd}                        }
    \sum_{                        \gamma_{ac} \gamma_{bd}}
(a b \alpha \beta  \vert \rho_{ab} c_{ab} \gamma_{ab}) 
(c d \gamma \delta \vert \rho_{cd} c_{cd} \gamma_{cd}) 
(c_{ab} c_{cd} \gamma_{ab} \gamma_{cd} \vert \rho e \varepsilon)
		\label{eq32} \\
&\times& 
(a c \alpha \gamma \vert \rho_{ac} c_{ac} \gamma_{ac})^* 
(b d \beta \delta \vert \rho_{bd} c_{bd} \gamma_{bd})^* 
(c_{ac} c_{bd} \gamma_{ac} \gamma_{bd} \vert \rho' e' \varepsilon')^*
		\nonumber 
		\end{eqnarray}
The introduction in these definitions of (\ref{eq13}) and the use of the great
orthogonality theorem for $G$ leads to the properties
\begin{eqnarray}
& & (a(bc)\rho_{bc}c_{bc} \rho' d' \delta' \vert (ab) \rho_{ab}c_{ab} c \rho d \delta) 
\nonumber \\
&=& \delta_{d d'} \delta_{\delta \delta'} [d]^{-1} \sum_{\delta} 
(a(bc) \rho_{bc} c_{bc} \rho' d \delta \vert (ab) 
\rho_{ab}c_{ab} c \rho d \delta)
\label{eq33}
\end{eqnarray}
and
\begin{eqnarray}
& & ((ac)\rho_{ac}c_{ac} (bd)\rho_{bd}c_{bd}  \rho' e' \varepsilon' \vert 
(ab)\rho_{ab} c_{ab} (cd) \rho_{cd}c_{cd} \rho  e  \varepsilon )
\nonumber \\
& & = \delta_{e e'} \delta_{\varepsilon \varepsilon'} [e]^{-1} \sum_{\varepsilon}
((ac) \rho_{ac} c_{ac} (bd) \rho_{bd} c_{bd} \rho' e \varepsilon \vert 
 (ab) \rho_{ab} c_{ab} (cd) \rho_{cd} c_{cd} \rho  e \varepsilon)
\label{eq34}
\end{eqnarray}
so that the recoupling coefficients defined by (\ref{eq31}) and (\ref{eq32}) are 
basis-independent (i.e., they do not depend on the labels of type $\alpha$) in
contrast with the coupling coefficients $(a b \alpha \beta \vert \rho c \gamma)$. 

By using the orthonormality of the CGCs, it can be shown that the CCGs occurring in 
Eqs.~(\ref{eq31}) and (\ref{eq32}) can be moved from the right hand side to the left 
hand side in such a way to produce new relations for which the total number of CGCs 
remains equal to 4 and 6, respectively. Repeated actions of this type lead to 
orthonormality relations for the recoupling coefficients (\ref{eq31}) and (\ref{eq32}). 

In a way paralleling the passage from the coupling coefficients to the 
$($3--$a \alpha)_\rho$ symbol, one can define $($6--$a)_{4\rho}$ and $($9--$a)_{6\rho}$ symbols 
from the recoupling coefficients defined by (\ref{eq31})-(\ref{eq34}). The defining 
expressions $($6--$a)_{4\rho}$ and $($9--$a)_{6\rho}$ symbols are very complicated and not
especially instructive in the case of an arbitrary compact group $G$. Hence, 
they shall be omitted as well as the defining expressions for higher
$(3N$--$a)_{2N\rho}$ symbols corresponding to the recoupling of $N \geq 4$
IRCs. Finally, note that the recoupling coefficients and their associated 
$(3N$--$a)_{2N\rho}$ symbols,  $N > 1$,  for a $G$ group can be connected to  
other basis-independent quantities, viz., the characters of $G$ \cite{WTSharp,KibJPhysA}.

\subsection{Irreducible tensorial sets}

Let $\lbrace \vert \tau a \alpha) : \alpha = 1,2,\ldots,[a] \rbrace$ 
be a basis for the irreducible matrix representation $D^a$ of $G$. 
The vectors $\vert \tau a \alpha)$ are defined on a unitary or pre-Hilbert 
space $\cal E$ (indeed, a Hilbert space in the quantum-mechanical applications)
and there exists an application $R \mapsto P_R$ such that
\begin{eqnarray}
P_R \vert \tau a \alpha ) = \sum_{\alpha'=1}^{[a]} 
    \vert \tau a \alpha') D^a(R)_{\alpha' \alpha}
\label{eq35}
\end{eqnarray}
for any $R$ in $G$. Following the work by Fano and Racah 
\cite{FanoRacah} on the $SU(2)$ group, we refer the set 
$\lbrace \vert \tau a \alpha) : \alpha = 1,2,\ldots,[a] \rbrace$ 
to as an irreducible tensorial set (ITS) of vectors
associated with $D^a$. The label $\tau$ may serve to distinguish 
different ITSs of vectors associated with the same irreducible matrix
representation $D^a$. (In practical applications, this label consists of
various quantum numbers arising from nuclear, or atomic or molecular 
configurations.) In this connection, note the following standardization: It is always 
possible to arrange that 
$\lbrace \vert \tau a \alpha) : \alpha = 1,2,\ldots,[a] \rbrace$ and
$\lbrace \vert \tau'a \alpha) : \alpha = 1,2,\ldots,[a] \rbrace$ span the same 
matrix representation $D^a$ rather than two equivalent representations. We 
shall assume that such a standardization is always satisfied.

From two ITSs
$\lbrace \vert \tau_a a \alpha) : \alpha = 1,2,\ldots,[a] \rbrace$  and 
$\lbrace \vert \tau_b b \beta ) : \beta  = 1,2,\ldots,[b] \rbrace$, we can 
construct another ITS of vectors. Let us define
		\begin{eqnarray}
\vert \tau_a \tau_b a b \rho c \gamma) := \sum_{\alpha \beta} 
\vert \tau_a a \alpha) \otimes 
\vert \tau_b b \beta ) (a b \alpha \beta \vert \rho c \gamma)
		\label{eq36}
		\end{eqnarray}
Then, as a simple corollary of (\ref{eq7}), the set 
$\lbrace \vert \tau_a\tau_bab \rho c \gamma) : \gamma = 1,2,\ldots,[c] \rbrace$ 
can be shown to be an ITS associated with $D^c$. 

In a similar way, let us consider a set 
$\lbrace T_\alpha^a : \alpha = 1,2,\ldots,[a] \rbrace$ 
of (linear) operators defined on $\cal E$ and such that
\begin{eqnarray}
P_R T_{\alpha }^a P_R^{-1} = \sum_{\alpha'=1}^{[a]} 
    T_{\alpha'}^a D^a(R)_{\alpha'\alpha}
\label{eq37}
\end{eqnarray}
for any $R$ in $G$. This set is called an ITS of operators associated with
$D^a$. We also say that this set defines an irreducible tensor operator
${\bf T}^a$ associated with $D^a$. 
Note the implicit standardization: The sets
$\lbrace T_\alpha^a : \alpha = 1,2,\ldots,[a] \rbrace$ and 
$\lbrace U_\alpha^a : \alpha = 1,2,\ldots,[a] \rbrace$ span the same matrix 
representation $D^a$ rather than two equivalent representations.

In full analogy with (\ref{eq36}), we define
		\begin{eqnarray}
\lbrace {\bf T}^a \otimes {\bf U}^b\rbrace^{\rho c}_\gamma := \sum_{\alpha \beta}
T_\alpha^a U_\beta^b (a b \alpha \beta \vert \rho c \gamma)
		\label{eq38}
		\end{eqnarray}
from the two ITSs 
$\lbrace T_\alpha^a : \alpha = 1,2,\ldots,[a] \rbrace$ and 
$\lbrace U_\alpha^b : \beta  = 1,2,\ldots,[b] \rbrace$. As a result, the set
$\lbrace \lbrace {\bf T}^a \otimes 
                 {\bf U}^b \rbrace^{\rho c}_\gamma : \gamma = 1,2,\ldots,[c] \rbrace$ 
is an ITS of operators associated with $D^c$. We say that
$\lbrace {\bf T}^a \otimes {\bf U}^b \rbrace$ is the direct product of the 
irreducible 
tensor operators ${\bf T}^a$ and ${\bf U}^b$. 
Observe that this direct product defines a 
tensor operator which is reducible in general. Equation (\ref{eq38}) gives the various
irreducible components of $\lbrace {\bf T}^a \otimes {\bf U}^b \rbrace$.

\subsection{The Wigner-Eckart theorem}

The connection between most of the quantities introduced up to now appears in
the calculation of the matrix element 
$(\tau'a'\alpha'\vert T^b_{\beta} \vert \tau a \alpha)$, the scalar 
product on $\cal E$ of the 
                     $T^b_{\beta} \vert \tau  a  \alpha) $ vector by the  
                                 $\vert \tau' a' \alpha')$ vector. By developing 
the identity 
\begin{eqnarray}
(\tau' a' \alpha' \vert T^b_{\beta} \vert 
 \tau  a  \alpha) = 
(\tau' a' \alpha' \vert P_R^{\dagger} P_R 
          T^b_{\beta}   P_R^{     -1} P_R \vert 
 \tau  a  \alpha)
\label{eq39}
\end{eqnarray}
we get, after some manipulations, the following basic theorem.

{\bf Theorem 1} (Wigner-Eckart's theorem). The scalar product 
$(\tau' a' \alpha' \vert T^b_\beta \vert \tau a \alpha)$ can be decomposed as 
		\begin{eqnarray}
(\tau' a' \alpha' \vert T_\beta^b \vert \tau a \alpha) = \sum_{\rho}
( \tau' a' \vert\vert T^b \vert\vert \tau a )_\rho \; 
\sum_{ a'' \alpha'' } \pmatrix{
a''&a'\cr
\alpha''&\alpha'\cr
}
\pmatrix{
b&a&a''\cr
\beta&\alpha&\alpha''\cr
}^*_\rho
		\label{eq40a}
		\end{eqnarray}
Alternatively, (\ref{eq40a}) can be cast into the form
		\begin{eqnarray}
(\tau' a' \alpha' \vert T_\beta^b \vert \tau a \alpha) = [a']^{-\frac{1}{2}} 
\sum_{\rho } 
\langle \tau' a' \vert\vert T^b \vert\vert \tau a \rangle_\rho
 (a b \alpha \beta \vert \rho a' \alpha')^*
		\label{eq40b}
		\end{eqnarray}
with 
		\begin{eqnarray}
\langle \tau' a' \vert\vert T^b \vert\vert \tau a \rangle_\rho := 
                              \sum_{\rho'} M (ab,a')_{\rho \rho'}^*
      ( \tau' a' \vert\vert T^b \vert\vert \tau a       )_{\rho'}
		\label{40ab}
		\end{eqnarray}
where $M(ab,a')$ is an arbitrary unitary matrix (cf., (\ref{eq24}) and (\ref{eq25})).

In the summation-factorization afforded by (\ref{eq40a}) or (\ref{eq40b}), 
there are two types of terms, namely, the $($3--$a \alpha)_\rho$ symbols or 
the CGCs $(a b \alpha \beta \vert \rho a' \alpha')$ that depend on the
$G$ group only and the so-called reduced matrix elements 
$      ( \tau' a' \vert\vert T^b \vert\vert \tau a       )_\rho$ or 
$\langle \tau' a' \vert\vert T^b \vert\vert \tau a \rangle_\rho$  
that depend both on $G$ and on the physics of the problem under consideration. 
The reduced matrix elements do not depend on the `magnetic quantum numbers' 
($\alpha'$, $\beta$ and $\alpha$) and therefore, like the
recoupling coefficients, are basis-independent. We then understand the interest
of the recoupling coefficients in applications: The reduced matrix elements 
for a composed system may be developed as functions of reduced matrix elements 
for elementary systems and recoupling coefficients. In this direction, it can be
verified that the matrix element
$(\tau_a'\tau_b'a'b'\rho'c'\gamma'\vert\lbrace {\bf T}^d \otimes {\bf U}^e\rbrace{\sigma f
\atop \varphi} \vert \tau_a\tau_b a b\rho c \gamma)$ 
can be expressed in terms
of the recoupling coefficients defined by (\ref{eq32}) and (\ref{eq34}). 

Equations (\ref{eq40a}) and (\ref{eq40b}) generalize the Wigner-Eckart theorem originally 
derived by Eckart for vector operators of the rotation group \cite{eckart30}, by Wigner 
for tensor operators of the rotation group \cite{wigner31} and of simply reducible groups 
\cite{wigner65}, and by Racah for tensor operators of the rotation group \cite{RII}. 

A useful selection rule on the matrix element 
$(\tau' a' \alpha' \vert T^b_\beta \vert \tau a \alpha)$ immediately follows
from the CGCs in (\ref{eq40b}). The latter matrix element vanishes if the
direct product $a \otimes b$ does not contains $a'$. Consequently, in order to
have $(\tau' a' \alpha' \vert T^b_\beta \vert \tau a \alpha) \ne 0$, it is
necessary (but not sufficient in general) that the IRC $a'$ be contained in 
$a \otimes b$. 

As an interesting particular case, let us consider the situation where $b$ 
is the identity IRC of $G$. This means that the operator $H=T^0_0$ is 
invariant under $G$ (see (\ref{eq37})). Equation (\ref{eq40b}) can be particularized to    
\begin{eqnarray}
(\tau' a' \alpha' \vert H \vert \tau a \alpha) = 
\delta_{a a'} \delta_{\alpha \alpha'} 
\langle \tau' a \vert\vert T^0 \vert\vert \tau a \rangle 
\label{eq41}
\end{eqnarray}
where the index $\rho$ is not necessary since $a \otimes 0 = a$. The Kronecker
deltas in (\ref{eq41}) show that there are no $a'$-$a$ and/or $\alpha'$-$\alpha$
mixing. We say that $a$ and $\alpha$ are `good quantum numbers' 
for $H$. The initial and final states have the same quantum numbers as far as
these numbers are associated with the invariance group $G$. 
The invariant $H$ does not mix state vectors belonging to different 
irreducible representations $a$ and $a'$. 
Furthermore, it does not mix state vectors 
belonging to the same irreducible representation $a$ but having different 
labels $\alpha$ and $\alpha'$. 

It is very important to realize that phase factors of type $(-1)^a$, 
$(-1)^{a - \alpha}$ and $(-1)^{a+b+c}$ do not appear in (\ref{eq40a}) 
and (\ref{eq40b}). Indeed, the present exposure is entirely free of
such phase factors, in contrast with other presentations. As a matter of fact, in many
works the passage from the Clebsh-Gordan or unsymmetrical form to the
$($3--$a \alpha)_\rho$ or symmetrical form of the coupling coefficients involves
unpleasant questions of phase. This is not the case in (\ref{eq24}) and (\ref{eq25}). Such
a fact does not mean that (\ref{eq24}) and (\ref{eq25}) as well as other general
relations are free of arbitrary phase factors. In fact, all the phase
factors are implicitly contained in the matrices $M$, the 2--$a \alpha$ 
symbols and the (basis-independent) Frobenius-Schur coefficient.

\subsection{The Racah lemma}

We have already emphasized the interest of considering chains of groups
rather than isolated groups. Let us now denote $G$ as $G_{a}$ and let 
$G_{\Gamma}$ be a subgroup of $G_{a}$. In this case, the labels 
of type $\alpha$, that occur in what precedes, may be replaced by triplets of
type $\alpha \Gamma \gamma$. The label of type $\Gamma$ stands for an IRC of
the group $G_{\Gamma}$, the label of type $\gamma$ is absolutely necessary when 
$[{\Gamma}] > 1$ and the new label
of type $\alpha$ is a branching multiplicity label to be used when the $\Gamma$ 
IRC of $G_{\Gamma}$ is contained several times in the $a$ IRC of the $G_{a}$ head group. (The 
$\gamma$ label is an internal multiplicity label for $G_{\Gamma}$ and the $a$ label 
is an external multiplicity label inherent to the restriction $G_{a} \to G_{\Gamma}$.) Then, 
the $( a_1 a_2 \alpha_1 \alpha_2 \vert \rho a \alpha )$ CGC for the $G_{a}$ group is replaced by the 
 $( a_1 a_2 \alpha_1 \Gamma_1 \gamma_1 
            \alpha_2 \Gamma_2 \gamma_2 \vert \rho a \alpha \Gamma \gamma )$ CGC for
the $G_{a}$ group in a $G_{a} \supset G_{\Gamma}$ basis. We can prove the following theorem. 

{\bf Theorem 2} (Racah's lemma). The CGCs of the $G_{a}$ group 
in a $G_{a} \supset G_{\Gamma}$ basis can be developed according to 
\begin{eqnarray}
( a_1 a_2 \alpha_1 \Gamma_1 \gamma_1 
          \alpha_2 \Gamma_2 \gamma_2 \vert \rho a \alpha \Gamma \gamma ) =
\sum_{\beta} 
( \Gamma_1 \Gamma_2 \gamma_1 \gamma_2 \vert \beta \Gamma \gamma )
(a_1 \alpha_1 \Gamma_1 + 
 a_2 \alpha_2 \Gamma_2 \vert \rho a \alpha \Gamma)_\beta 
\label{eq40c}
\end{eqnarray}
where the $( \Gamma_1 \Gamma_2 \gamma_1 \gamma_2 \vert \beta \Gamma \gamma )$ coefficients 
are CGCs for the $G_{\Gamma}$ group considered as an isolated group and the 
$(a_1 \alpha_1 \Gamma_1 + 
  a_2 \alpha_2 \Gamma_2 \vert \rho a \alpha \Gamma)_\beta$ coefficients do not depend on 
$\gamma_1$, $\gamma_2$ and $\gamma$. 

The proof of Racah's lemma was originally obtained from Schur's lemma \cite{RIV}. However, the 
analogy between (\ref{eq40a}), (\ref{eq40b}) and (\ref{eq40c}) should be noted. Hence, the Racah lemma
for a $G_{a} \supset G_{\Gamma}$ chain may be derived from the Wigner-Eckart theorem, for the $G_{a}$ 
group in a $G_{a} \supset G_{\Gamma}$ basis, applied to the Wigner operator, i.e., the operator 
whose matrix elements are the CGCs. The 
$(a_1 \alpha_1 \Gamma_1 + 
 a_2 \alpha_2 \Gamma_2 \vert \rho a \alpha \Gamma)_\beta$ 
in the development given by (\ref{eq40c}) are sometimes named isoscalar factors, 
a terminology that comes from the $SU(3) \supset U(1) \otimes SU(2)$ 
chain used in the eightfold way model of subatomic physics. 

From a purely group-theoretical point of view, it is worth to note that Racah's
lemma enables us to calculate the CGCs of the $G_{\Gamma}$ subgroup of $G_{a}$ 
when those of the $G_{a}$ group are known (see for example \cite{kiblerSweden} 
and references therein). In particular, for those triplets 
$(\Gamma_1\Gamma_2\Gamma)$ for which $\Gamma_1 \otimes \Gamma_2$ contains
$\Gamma$ only once, the CGCs 
$( \Gamma_1 \Gamma_2 \gamma_1 \gamma_2 \vert \Gamma \gamma )$ are given by a
simple formula in terms of the CGCs of $G_{a}$. 

The summation-factorization in (\ref{eq40c}) can be applied to each CGC 
entering the definition of any recoupling coefficient for the $G_{a}$ group. 
Therefore, the recoupling coefficients for $G_{a}$ can be developed in terms 
of the recoupling coefficients for its subgroup $G_{\Gamma}$ \cite{CRAS,kibler79}.

\subsection{Illustrative examples}

\subsubsection{The $SU(2)$ group in a $SU(2) \supset U(1)$ basis}

As a first example, we take $G_{a} \equiv SU(2)$ and $G_{\Gamma} \equiv U(1)$ where $SU(2$ and $U(1)$ 
are the universal covering groups or, in the terminology of molecular physics, the `double' groups 
of the proper rotation groups $R(3) \sim SO(3)$ and $R(2) \sim SO(2)$, respectively. In
this case, $a \equiv (j)$ where $j$ is either an integer (for vector representations) or 
a half-of-an-odd integer (for spinor representations), $\alpha \Gamma \gamma \equiv m$ 
ranges from $-j$ to $j$ by unit step, and $D^a(R)_{\alpha\alpha'}$ can be identified to 
the element $D^{(j)}(R)_{mm'}$ of the well-known Wigner rotation matrix of
dimension $[j] \equiv 2j +1$. The matrix representation $D^{(j)}$
corresponds to the standard basis 
$\lbrace \vert j , m) : m = j, j - 1, \ldots, -j \rbrace$ where
$\vert j , m)$ denotes an eigenvector of the (generalized) 
angular momentum operators $J^2$ and
$J_z$. (For $j$ integer, the label $\ell$ often replaces $j$.) The labels of 
type $m$
clearly refer to IRCs of the rotation group $C_{\infty} \sim R(2)$. 
Therefore, the basis 
$\lbrace \vert j , m) : m = j, j - 1, \ldots, -j \rbrace$
is called a $R(3) \supset R(2)$ or $SU(2) \supset U(1)$ basis. Furthermore, the
multiplicity label $\rho$ is not necessary since $SU(2)$ is multiplicity-free.
Consequently, the (real) CGCs of $SU(2)$ in a $SU(2) \supset U(1)$ 
basis are written $(j_1j_2m_1m_2\vert jm)$. They are also called Wigner 
coefficients. 

In view of the ambivalent nature of $SU(2)$, the 2--$a\alpha$ symbol reduces here to 
		\begin{eqnarray}
\pmatrix{
j&j'\cr
m&m'\cr
} = \delta_{jj'}
\pmatrix{
    j     \cr
m \quad m'\cr
}
		\label{eq42}
		\end{eqnarray}
We can take 
		\begin{eqnarray}
\pmatrix{
     j     \cr
m  \quad m'\cr
}
:= (-1)^{j+m} \delta (m',-m)
		\label{eq42bis}
		\end{eqnarray}
where $(-1)^{j+m} \delta (m',-m)$ is a component of the  
1$-jm$ Herring-Wigner metric tensor (in the Edmonds normalization \cite{Edmonds}). Then, the
introduction of (\ref{eq42}) and (\ref{eq42bis}) into (\ref{eq24}) for the $SU(2) \supset U(1)$ 
chain shows that the 3--$a\alpha$ symbol identifies to the 3--$jm$ Wigner symbol
		\begin{eqnarray}
\pmatrix{
j_1&j_2&j_3\cr
m_1&m_2&m_3\cr
} := (2j_3 + 1)^{-\frac{1}{2}} (-1)^{j_3 - m_3 - 2 j_2} 
( j_2 j_1 m_2 m_1 \vert j_3, - m_3 )
		\label{(43)} 
		\end{eqnarray} 
provided we chose $M(j_2j_1,j_3) = (-1)^{2j_1}$. Such a choice ensures that 
the 3--$jm$ symbol is highly symmetrical under permutation of its columns.

In the $SU(2)$ case, the (6--$a$)$_{4\rho}$ and (9--$a$)$_{6\rho}$ 
symbols may be chosen to coincide  with  the 6--$j$ Wigner 
(or $\bar W$ Fano-Racah) symbol      and the 9--$j$ Wigner 
(or $     X$ Fano-Racah) symbol, respectively. More precisely, we have 
		\begin{eqnarray}
\left\lbrace\matrix{
j_1&j_{23}&j  \cr 
j_3&j_{12}&j_2\cr
}\right\rbrace
& := & (-1)^{j_1+j_2+j_3+j} [(2j_{12}+1) (2j_{23}+1)]^{-\frac{1}{2}}
		\nonumber \\
& \times & (j_1(j_2j_3)j_{23}jm \vert (j_1j_2)j_{12}j_3jm)
		\label{(eq44)}
		\end{eqnarray}
and 
		\begin{eqnarray}
\left\lbrace\matrix{
j_1   &j_2   &j_{12}\cr
j_3   &j_4   &j_{34}\cr
j_{13}&j_{24}&j     \cr
}\right\rbrace
& := & [(2j_{12} + 1)(2j_{34} + 1)(2j_{13} + 1)(2j_{24} + 1)]^{-\frac{1}{2}}
		\nonumber \\
& \times & ((j_1j_3)j_{13}(j_2j_4)j_{24}jm \vert (j_1j_2)j_{12}(j_3j_4)j_{34}jm)
		\label{(eq45)}
		\end{eqnarray}
in terms of recoupling coefficients (cf., (\ref{eq33}) and (\ref{eq34})).

Finally, for $a \equiv (k)$, the {\bf T}$^a$ ITS coincides with the {\bf T}$^{(k)}$ 
irreducible tensor operator of rank $k$ (and having $2 k + 1$ components) introduced 
by Racah. We denote by $T^{(k)}_q$ the components of {\bf T}$^{(k)}$ in a 
$SU(2) \supset U(1)$ basis.

All the relations of subsections 3.1-3.7 
may be rewritten as familiar relations of angular 
momentum theory owing to the just described correspondence 
rules. For example, (\ref{eq17}) or (\ref{eq18}) and (\ref{eq40a}) can 
be specialized to 
		\begin{eqnarray}
D^{(j)}(R)_{mm'}^* = (-1)^{m-m'} D^{(j)}(R)_{-m,-m'}
		\label{(46)}
		\end{eqnarray}
and
		\begin{eqnarray}
(\tau' j' m' \vert T^{(k)}_q \vert \tau j m) = (-1)^{j'-m'} \pmatrix{
j'&k&j\cr
-m'&q&m\cr
}
(\tau' j' \vert\vert T^{(k)} \vert\vert \tau j)
		\label{(47)}
		\end{eqnarray}
respectively. For more details, the reader should consult the textbooks in Refs. \cite{FanoRacah,Edmonds} (see 
also \cite{Juddbook, Wybournebook, Condon}).

\subsubsection{The $SU(2)$ group in a $SU(2) \supset G^*$ basis} 

We now consider the case $G_a \equiv SU(2)$ and $G_{\Gamma} \equiv G^*$, where $G^*$ is isomorphic to  
the double group of a point (proper) rotation group $G$. Then, we have $a \equiv (j)$ and we take 
$\alpha \Gamma \gamma \equiv a \Gamma \gamma$ for the labels $a$ and $\alpha \Gamma \gamma$ of 
Section 3.8. This will be clarified below. 

{\bf 1 - }{\bf The restriction of $SU(2)$ to $G^*$}

Each IRC ($j$) of $SU(2)$ can be decomposed into a direct sum of IRC's of $G^*$:
		\begin{eqnarray}
(j) = \sum_{\Gamma} \sigma(\Gamma \vert j) \Gamma
		\end{eqnarray}
where
		\begin{eqnarray}
\sigma(\Gamma \vert j) = \vert G^* \vert ^{-1} \int_{G^*} dR \chi^{\Gamma}(R)^* \chi^{(j)}(R)
		\end{eqnarray}
stands for the multiplicity of the $\Gamma$ IRC of $G^*$ in ($j$). In terms of unitary 
matrix representations, this means that
		\begin{eqnarray}
D^{(j)} \sim \bigoplus_{\Gamma} \sigma(\Gamma \vert j) D^{\Gamma}
		\end{eqnarray}
In other words, there exists a unitary matrix ${U^j}$ such that
		\begin{eqnarray}
({U^j})^\dagger D^{(j)}(R) {U^j} = \bigoplus_{\Gamma} \sigma(\Gamma \vert j) D^{\Gamma}(R)
		\end{eqnarray}
holds for any $R$ in $G^*$. This leads to
		\begin{eqnarray}
\sum_{mm'} (jm \vert j a  \Gamma  \gamma )^* D^{(j)}(R)_{mm'} 
           (jm'\vert j a' \Gamma' \gamma') = \delta_{aa'} \delta_{\Gamma\Gamma'} D^{\Gamma}(R)_{\gamma\gamma'}
		\label{52}
		\end{eqnarray}
or
		\begin{eqnarray}
D^{(j)}(R)_{mm'} = \sum_{a \Gamma \gamma \gamma'} (jm \vert j a \Gamma \gamma) 
                      D^{\Gamma}(R)_{\gamma\gamma'} (jm'\vert j a \Gamma \gamma')^*
		\label{53}
		\end{eqnarray}
for any $R$ in $G^*$. In (\ref{52}) and (\ref{53}), $(jm \vert ja \Gamma \gamma)$ denotes an
element of the matrix ${U^j}$:
		\begin{eqnarray}
(jm \vert j a \Gamma \gamma) := ({U^j})_{m,a\Gamma\gamma}
		\label{54}
		\end{eqnarray}
The label $a$ (cf., the column index $a \Gamma \gamma$ of ${U^j}$) is a branching
multiplicity label indispensable when $\Gamma$ appears more than once in ($j$). Note
that the unitary property of the matrix ${U^j}$ corresponds to $R = E$, the unit
element of $G^*$, in (\ref{52}) and (\ref{53}):
		\begin{eqnarray}
\sum_{m} (j m \vert j a \Gamma \gamma)^* (j m \vert j a' \Gamma' \gamma') =
\delta_{aa'} \delta_{\Gamma\Gamma'} \delta_{\gamma\gamma'}
		\end{eqnarray}
or inversely 
		\begin{eqnarray}
\sum_{a \Gamma \gamma} (j m \vert ja \Gamma \gamma) (j m' \vert ja \Gamma \gamma)^* = \delta_{mm'}
		\end{eqnarray}
Observe that (\ref{52}) and (\ref{53}) are note sufficient for determining the
reduction coefficients $(jm \vert j a \Gamma \gamma)$ once the irreducible
representation matrices of $G^*$ and $SU(2)$ are known since the coefficients
		\begin{eqnarray}
(j m \vert j b \Gamma \gamma) := \sum_{a} 
(j m \vert j a \Gamma \gamma) {M}_{ab}
		\end{eqnarray}
where ${M}$ is an arbitrary unitary matrix satisfy (\ref{52}) and (\ref{53}) with the
replacement $a \rightarrow b$. Nevertheless, (\ref{52}) and (\ref{53}) lead to systems
that may be useful for the calculation of the $(j m \vert j a \Gamma \gamma)$ coefficients.

{\bf 2 - }{\bf Irreducible tensorial sets}

From the ITS of vectors $\lbrace \vert \tau j m) : m = j, j-1, \ldots, -j \rbrace$
associated with $D^{(j)}$, we define
		\begin{eqnarray}
\vert \tau j a \Gamma \gamma) := \sum_{m} \vert \tau j m) (j m \vert j a \Gamma \gamma)
		\label{58}
		\end{eqnarray}
Equation (\ref{53}) allows us to show
		\begin{eqnarray}
P_R \vert \tau j a \Gamma \gamma) = \sum_{\gamma'} \vert \tau j a \Gamma \gamma') D^{\Gamma}(R)_{\gamma'\gamma}
		\end{eqnarray}
for any $R$ in $G^*$. Similarly, from the ITS of operators $\lbrace T_q^{(k)} : q = k, k-1, \ldots, -k \rbrace$ 
associated with ${\bf D}^{(k)}$, we define
		\begin{eqnarray}
T_{\gamma}^{(k a \Gamma)} \equiv T_{a \Gamma \gamma}^{(k)} := \sum_{q} T_q^{(k)} (kq \vert k a \Gamma \gamma)
		\label{60}
		\end{eqnarray}
so that
		\begin{eqnarray}
P_R T_{\gamma}^{(k a \Gamma)} P_R^{-1} = \sum_{\gamma'} T_{\gamma'}^{(k a \Gamma)} D^{\Gamma}(R)_{\gamma'\gamma}
		\end{eqnarray}
holds for any $R$ in $G^*$.

At this point, it is important to remark that (\ref{58}) and (\ref{60}) provide us
with ITSs both for $SU(2)$ and $G^*$. Indeed $\lbrace \vert \tau j a \Gamma \gamma) : \gamma \ {\rm ranging} \ \rbrace$ 
is an ITS of vectors spanning the matrix representation $D^{\Gamma}$ of $G^*$ while 
$\lbrace \vert \tau j a \Gamma \gamma) : a \Gamma \gamma \ {\rm ranging} \ \rbrace$ is 
an ITS of vectors spanning the matrix representation ${\cal D}^{(j)}$ of $SU(2)$ defined by
		\begin{eqnarray}
{\cal D}^{(j)}(R)_{a \Gamma \gamma,a' \Gamma' \gamma'} := \sum_{mm'} (jm \vert j a \Gamma \gamma)^* 
D^{(j)}(R)_{mm'} (jm' \vert j a' \Gamma' \gamma')
		\end{eqnarray}
for any $R$ in $SU(2)$. A similar remark applies to the sets 
$\lbrace T_{\gamma}^{(ka\Gamma)} : \gamma \ {\rm ranging} \ \rbrace$ and 
$\lbrace T_{a \Gamma \gamma}^{(k)} : a \Gamma \gamma \ {\rm ranging} \rbrace$.

{\bf 3 - }{\bf Wigner-Eckart theorems}

As an important consequence of the latter two remarks, we may apply the
Wigner-Eckart theorem either for the group $SU(2)$ in a $SU(2) \supset G^*$ basis or 
for the group $G^*$ in a $G^* \subset SU(2)$ basis. For $G^*$ in a $G^* \subset SU(2)$
basis, (\ref{eq40a}) gives
		\begin{eqnarray}
(\tau_1j_1a_1\Gamma_1\gamma_1\vert T^{(ka\Gamma)}_\gamma\vert
 \tau_2j_2a_2\Gamma_2\gamma_2) &=& \sum_{\rho}
(\tau_1j_1a_1\Gamma_1\vert\vert T^{(ka\Gamma)}\vert\vert
 \tau_2j_2a_2\Gamma_2)_\rho  
\nonumber \\
& \times & \sum_{\Gamma'_1\gamma'_1} \pmatrix{
\Gamma_1'&\Gamma_1\cr
\gamma_1'&\gamma_1\cr
}
\pmatrix{
\Gamma & \Gamma_2 & \Gamma_1'\cr
\gamma & \gamma_2 & \gamma_1'\cr
}_\rho^*
		\label{63}
		\end{eqnarray}
For $SU(2)$ in a $SU(2) \supset G^*$ basis, we can combine (\ref{eq40b}), (\ref{58}) and
(\ref{60}) to obtain the compact formula
		\begin{eqnarray}
(\tau_1j_1a_1\Gamma_1\gamma_1\vert T_{a \Gamma \gamma}^{(k)} \vert
\tau_2j_2a_2\Gamma_2\gamma_2) = (\tau_1j_1\vert\vert T^{(k)}\vert\vert \tau_2j_2)f
\pmatrix{
j_1 & j_2 & k\cr
a_1\Gamma_1\gamma_1 & a_2\Gamma_2\gamma_2 & a\Gamma \gamma\cr}
    \label{64}
		\end{eqnarray}
where the $f$ symbol is defined by 
		\begin{eqnarray}
f\pmatrix{
j_1&j_2&k\cr
a_1\Gamma_1\gamma_1&a_2\Gamma_2\gamma_2&a\Gamma \gamma\cr
} := (-1)^{2k} (2j_1+1)^{-1/2} (j_2ka_2\Gamma_2\gamma_2 a\Gamma \gamma\vert
j_1a_1\Gamma_1\gamma_1)^*
		\label{65}
		\end{eqnarray}
in function of the CGC
		\begin{eqnarray}
(j_1j_2a_1\Gamma_1\gamma_1a_2\Gamma_2\gamma_2\vert ja \Gamma \gamma) &:=& \sum_{m_1m_2m} (j_1m_1\vert j_1a_1\Gamma_1\gamma_1)^*
\nonumber \\
& \times & (j_2m_2\vert j_2a_2\Gamma_2\gamma_2)^* (j_1j_2m_1m_2\vert jm)(jm\vert ja \Gamma
\gamma)
		\label{66}
		\end{eqnarray}
of $SU(2)$ in a $SU(2) \supset G^*$ basis \cite{JMS68}. 

There are many advantages to use (\ref{64}) rather than (\ref{63}). In (\ref{63}),
both the reduced matrix elements and the coupling coefficients (cf., the
2--$a\alpha$ and $($3--$a\alpha)_\rho$ symbols) depend of the symmetry group $G^*$.
Furthermore, the factorization offered by (\ref{63}) is not complete in view of
the summation over the multiplicity label $\rho$. On the other side, the matrix
element given by (\ref{64}) factorizes in two parts: a coupling coefficient
(cf., the $f$ symbol) for the $SU(2) \supset G^*$ chain and a reduced matrix element
which does not depend of the group $G^*$. This maximal factorization takes place
even in the case where $G^*$ is not multiplicity-free. The reduced matrix elements
in (\ref{64}) applied to complex systems either are obtainable from tables or can
be calculated from Racah's method in terms of recoupling coefficients of
$SU(2)$, coefficients of fractional parentage, and elementary reduced matrix
elements. The main calculation to do when dealing with (\ref{64}) most of the
time concerns the $f$ geometrical coefficient, a quantity which is independent
of the additional quantum numbers $\tau_1$ and $\tau_2$ and which remains
invariant when the tensor operator ${\bf T}^k$ is replaced by any tensor operator
${\bf U}^k$ of the same rank.

The calculation of the $f$ coefficients defined by (\ref{65}) and (\ref{66}) touches a
simple problem of symmetry adaptation. In fact, the determination of the
symmetry-adapted CGCs (\ref{66}) require the knowledge of the reduction coefficients
(\ref{54}). These reduction coefficients are the expansion coefficients of symmetry
adapted functions (cf.~(\ref{58})) or symmetry-adapted operators (cf.~(\ref{60})
so that their calculation may be achieved by numerous means (resolution of
linear systems like (\ref{52}) or (\ref{53}), projection operator techniques, \ldots).

{\bf 4 - }{\bf The $\bar f$ symbol}

Equation (\ref{65}) shows that the behavior of the $f$ symbol under the interchange 
of its first and second columns is not easy to describe. The $f$ symbol may be
symmetrized owing to the introduction of the 1--$j a \Gamma \gamma$ symbol
		\begin{eqnarray}
\pmatrix{
                j                   \cr
a\Gamma\gamma \quad a'\Gamma'\gamma'\cr
}
:= \sum_{mm'} (jm\vert ja \Gamma \gamma)^* 
\pmatrix{
    j     \cr
m \quad m'\cr
}
(jm' \vert j a' \Gamma' \gamma')^*
		\end{eqnarray}
where the 1--$jm$ symbol is defined by (\ref{eq42}) and (\ref{eq42bis}). The $\bar f$ or 
3--$ja\Gamma\gamma$ symbol defined through
		\begin{eqnarray}
\bar f 
\pmatrix{j_1&j_2&j_3\cr
a_1\Gamma_1\gamma_1&a_2\Gamma_2\gamma_2&a_3\Gamma_3\gamma_3\cr
} &:=& \sum_{a_4 \Gamma_4 \gamma_4} 
\pmatrix{
                       j_3                       \cr
a_3 \Gamma_3 \gamma_3 \quad a_4 \Gamma_4 \gamma_4\cr
}
		\nonumber	\\	
& \times &		
f
\pmatrix{j_3&j_2&j_1\cr
a_4\Gamma_4\gamma_4& a_2\Gamma_2\gamma_2& a_1\Gamma_1\gamma_1\cr
}^*
		\label{68}
		\end{eqnarray}
then exhibits a high (permutational) symmetry since a simple developement of
(\ref{68}) leads to
		\begin{eqnarray}
\bar f\pmatrix{
j_1&j_2&j_3\cr
a_1\Gamma_1\gamma_1&a_2\Gamma_2\gamma_2&a_3\Gamma_3\gamma_3\cr
} = \sum_{m_1 m_2 m_3} \pmatrix{
j_1 & j_2 & j_3 \cr
m_1 & m_2 & m_3 \cr
}
{\prod_{i=1}^3} (j_i m_i \vert j_i a_i \Gamma_i \gamma_i)^*
		\label{eq69}
		\end{eqnarray}
(see \cite{JMS68} for the properties -- selection rules, orthogonality relations, etc. -- of the $f$ and $\bar f$ symbols).

For $G^* \equiv U(1)$ the $\bar f$ symbol and the 1--$j a \Gamma \gamma$ symbol
reduce to the 3--$jm$ Wigner symbol and to the 1--$jm$ Herring-Wigner symbol,
respectively. The 1--$j a \Gamma \gamma$ and $\bar f$ symbols are thus
2--$a\alpha$ and 3--$a\alpha$ symbols as defined in Sec. II (with $a \rightarrow j$ 
and $\alpha \rightarrow a \Gamma \gamma$), respectively, for the group
$SU(2)$ in a $SU(2) \supset G^*$ basis. The properties (existence conditions,
selection rules, symmetry properties, orthogonality properties, \ldots) of the
$\bar f$ (and $f$) symbols can be deduced from the ones of the 3--$jm$ symbols and the 
${U^j}$ matrices and have been discussed at length elsewhere \cite{JMS68, IJQC69, CRAS}. Let 
us simply mention that, by applying Racah's lemma, the $\bar f$ symbol can be developed as a linear 
combination of $($3--$\Gamma \gamma)_\rho$ according to 
		\begin{eqnarray}
\bar f\pmatrix{
j_1&j_2&j_3\cr
a_1\Gamma_1\gamma_1&a_2\Gamma_2\gamma_2&a_3\Gamma_3\gamma_3\cr
} = \sum_{\rho} 
\bar f \left( \pmatrix{
j_1&j_2&j_3\cr
a_1\Gamma_1 & a_2\Gamma_2 & a_3\Gamma_3 \cr
} \right)_{\rho} 
\pmatrix{
\Gamma_1 & \Gamma_2 & \Gamma_3 \cr
\gamma_1 & \gamma_2 & \gamma_3 \cr
}_{\rho}
		\label{eq70}
		\end{eqnarray}
where the $\bar f((\ldots))$ reduced coefficient is independent of $\gamma_1$, $\gamma_2$ et $\gamma_3$.

We are now in a position to enunciate correspondence rules for passing from the Wigner-Racah algebra of 
$SU(2)$ in a $SU(2) \supset U(1)$ basis (i.e., in the $\{ jm \}$ scheme) to the Wigner-Racah algebra of 
$SU(2)$ in a $SU(2) \supset G^*$  basis (i.e., in the $\{ j a \Gamma \gamma \}$ scheme): All the $m$- 
or $q$-dependent quantities are replaced by the corresponding $a \Gamma \gamma$-dependent quantities while 
the basis-independent quantitities (like 6--$j$ and 9--$j$ symbols) are unchanged. More precisely, we have 
		\begin{eqnarray}
		D^{(j)}(R)_{mm'}	& \to & {\cal D}^{(j)}_{a \Gamma \gamma , a' \Gamma' \gamma' } \nonumber \\
		(j_1 j_2 m_1 m_2 | jm) & \to & (j_1 j_2 a_1\Gamma_1\gamma_1 a_2\Gamma_2\gamma_2 | j a \Gamma \gamma) \nonumber \\
		1{\rm -}jm \ {\rm symbol} \ & \to & 1{\rm -}a \Gamma \gamma \ {\rm symbol} \nonumber \\
		3{\rm -}jm \ {\rm symbol} \ & \to & \bar f \ {\rm symbol} \label{eq97} \\
		3(n-1){\rm -}j \ {\rm symbol} \ & \to & 3(n-1){\rm -}j \ {\rm symbol} \nonumber \\
		\vert \tau j m) & \to & \vert \tau j a \Gamma \gamma) \nonumber \\
		T^{(k)}_q & \to & T^{(k)}_{a \Gamma \gamma} \nonumber
		\end{eqnarray} 
(see \cite{IJQC69} for more details).

\subsubsection{The $G^*$ group in a $G^* \subset SU(2)$ basis} 

{\bf 1 - }{\bf The general case}

Equations (\ref{eq12})-(\ref{eq15}) were used in numerous works for calculating coupling 
coefficients and $V$ symbols of subgroups of $SU(2)$. (Following Griffith \cite{griffith62}, 
the $($3--$a \alpha)_\rho$ symbols of a group of molecular interest are referred to as $V$ 
symbols in what follows.) We now describe an alternative method for calculating the $V$ coefficients of 
a subgroup $G^*$ of $SU(2)$ as renormalized ${\bar f}$ coefficients of the $SU(2) \supset G^*$ chain. This 
method combines three basic ingredients scattered in various (implicit or explicit) approaches starting with 
the pioneer works by Tanabe, Sugano and Kamimura: the concept of quasi angular momentum, 
the definition of the ${\bar f}$ symbol and renormalization techniques. For the purpose of simplicity, we 
shall limit ourselves to a multiplicity-free group $G^*$ but it should be noted that the method may be extended 
to an arbitrary subgroup of $SU(2)$.

Given the $\Gamma$ IRC of $G^*$, let $({\hat {\jmath}}(\Gamma))$ or simply $({\hat{\jmath}})$ the IRC of $SU(2)$ that contains 
$\Gamma$ once and only once. Thus, ${\hat{\jmath}}$ is the smallest value of $j$ for which $\sigma(\Gamma | j) = 1$. The 
value ${\hat{\jmath}}$ refers to a quasi angular momentum \cite{Vilnius} (see \cite{kiblerSweden} too). In the 
multiplicity-free case where the identity IRC of $G^*$ appears once and only once in the triple direct product 
$\Gamma_1 \otimes \Gamma_2 \otimes \Gamma_3$, there is no need for the internal multiplicity label 
$\rho$ in the 3--$\Gamma \gamma$ or $V$ symbol. Therefore, let us put  
		\begin{eqnarray}
V \pmatrix{
\Gamma_1 & \Gamma_2 & \Gamma_3 \cr
\gamma_1 & \gamma_2 & \gamma_3 \cr
} &:=& x(\Gamma_1 \Gamma_2 \Gamma_3)
\bar f \pmatrix{
{\hat{\jmath}_1}      &{\hat{\jmath}_2}      &{\hat{\jmath}_3}      \cr
\Gamma_1\gamma_1&\Gamma_2\gamma_2&\Gamma_3\gamma_3\cr
} 
		\nonumber \\
& \times &
\left[ \sum_{\gamma_1 \gamma_2 \gamma_3} 
\left|  
\bar f\pmatrix{
{\hat{\jmath}_1}      &{\hat{\jmath}_2}      &{\hat{\jmath}_3}      \cr
\Gamma_1\gamma_1&\Gamma_2\gamma_2&\Gamma_3\gamma_3\cr
} 
\right|^2 \right]^{-1/2}
		\label{eq72}
		\end{eqnarray}
where $x(\Gamma_1 \Gamma_2 \Gamma_3)$ is an a	arbitrary phase factor that depends on 
$\Gamma_1$, $\Gamma_2$ and $\Gamma_3$ only. It can be verified by repeated application of 
(\ref{eq70}) that the $V$ symbol defined by (\ref{eq72}) satisfies (\ref{eq27})
and (\ref{eq26}) for $G^*$. Consequently, the $V$ symbol is nothing but a 
3--$\Gamma \gamma$ symbol for the $G^*$ group compatible with the choice implicitly assumed 
through (\ref{eq69}) with $j={\hat {\jmath}}(\Gamma)$ for the representation matrices 
$D^{\Gamma}$. 

For the sake of simplifying calculations with (\ref{eq72}), it should be noted that 
		\begin{eqnarray}
\sum_{\gamma_1 \gamma_2 \gamma_3} 
\left|  
\bar f \pmatrix{
{\hat{\jmath}_1}      &{\hat{\jmath}_2}      &{\hat{\jmath}_3}      \cr
\Gamma_1\gamma_1&\Gamma_2\gamma_2&\Gamma_3\gamma_3\cr
} 
\right|^2 = [\Gamma_i]
\sum_{{ \rm all \, } \gamma_k { \, \rm except \, } \gamma_i} 
\left|  
\bar f \pmatrix{
{\hat{\jmath}_1}      &{\hat{\jmath}_2}      &{\hat{\jmath}_3}      \cr
\Gamma_1\gamma_1&\Gamma_2\gamma_2&\Gamma_3\gamma_3\cr
} 
\right|^2  
		\label{eq73}
		\end{eqnarray}
for $i=1$, 2 or 3. In addition, if two of the three $\Gamma$'s are equivalent to 
two of the corresponding three $(\hat{\jmath})$'s, the right-hand side of (\ref{eq73})
can be simplified and (\ref{eq72}) takes a simple form. For instance, in the 
case $({\hat{\jmath}_1}) \equiv \Gamma_1$ and   
$({\hat{\jmath}_2}) \equiv \Gamma_2$, (\ref{eq72}) becomes
		\begin{eqnarray}
V \pmatrix{
\Gamma_1 & \Gamma_2 & \Gamma_3 \cr
\gamma_1 & \gamma_2 & \gamma_3 \cr
} = x(\Gamma_1 \Gamma_2 \Gamma_3) [\Gamma_3]^{- 1/2} (2 \hat{\jmath}_3 + 1)^{1/2}
\bar f \pmatrix{
{\hat{\jmath}_1} & {\hat{\jmath}_2} & {\hat{\jmath}_3} \cr
\Gamma_1\gamma_1 & \Gamma_2\gamma_2 & \Gamma_3\gamma_3 \cr
}
		\label{eq74}
		\end{eqnarray}
which is very simple to handle.

The main advantages of the method based on (\ref{eq72})-(\ref{eq74}) for 
calculating the $V$ coefficients of $G^*$ may be seen to be the following. First, 
the calculation is easy in the sense that the $V$ coefficients are deduced from 
a minimal set of $\bar f$ coefficients which are readily calculated (by hand or 
with the help of a computer) from (\ref{eq69}). The thus obtained $V$ coefficients 
of the $G^*$ group are simple linear combinations of 3--$jm$ coefficients for the 
$SU(2) \supset U(1)$ chain. Second, such a method allows us to work with bases of 
interest for molecular physics and quantum chemistry. In this respect, we may use 
in (\ref{eq69}) reduction coefficients $(jm|ja\Gamma\gamma)$ corresponding to 
Cartesian $p$, $d$ and $f$ spin-orbitals or corresponding to a chain of groups 
(for instance, the $SU(2) \supset O^* \supset D_4^* \supset D_2^*$ tetragonal 
chain or the $SU(2) \supset O^* \supset D_3^* \supset C_3^*$ trigonal chain). Third,
it is possible to transfer some of the features (formulas, symmetry properties, \ldots) 
of the 3--$jm$ symbol from the $SU(2) \supset U(1)$ standard chain to the $V$ symbol of 
$G^*$. For example, the permutation symmetry properties of the $V$ symbol can be 
chosen to be essentially the ones of the 3--$jm$ symbol. In fact, by choosing 
$x(\Gamma_1 \Gamma_2 \Gamma_3)$ invariant under the 3! permutations of its arguments, 
the $V$ symbol given by (\ref{eq72})-(\ref{eq74}) is multiplied by 
$(-1)^{{\hat{\jmath}}_1(\Gamma_1) + {\hat{\jmath}}_2(\Gamma_2) + {\hat{\jmath}}_3(\Gamma_3)}$ 
under an odd permutation of its columns so that is is invariant under an even permutation.

{\bf 2 - }{\bf Application to the octahedral group}

As an illustration, we consider the case where $G^*$ is the $O^*$ double octahedral group 
and limit ourselves to the determination of the $V$ coefficients of the $O$ octahedral 
group. Therefore, we can replace $SU(2) \supset O^*$ by $SO(3) \supset O$. The restriction of 
$SO(3)$ to $O$ yields
		\begin{eqnarray}
{\hat {\jmath}}(A_1) = 0, \quad 
{\hat {\jmath}}(A_2) = 3, \quad 
{\hat {\jmath}}(E  ) = 2, \quad 
{\hat {\jmath}}(T_1) = 1, \quad 
{\hat {\jmath}}(T_2) = 2 
		\label{eq75}
		\end{eqnarray}
where $A_1$, $A_2$, $E$, $T_1$ and $T_2$ denote the various IRCs of $O$.
In view of the permutation symmetry properties of the $V$ symbol, there are {\em a priori} 39 independent 
$V$ coefficients to be calculated for the $O$ group. The $\vert {\hat {\jmath}} \Gamma \gamma )$ vectors (the label 
$a$ is not necessary here) required for calculating these coefficients are given by 
		\begin{eqnarray}
&& \vert 0 A_1 a_1 ) 		=  \vert 0 , 0 )  																				\nonumber    \\
&& \vert 3 A_2 a_2 ) 		=  \frac{1}{\sqrt{2}} [ \vert 3 , 2 ) - \vert 3 , -2 ) ]  \nonumber    \\
&& \vert 2 E \theta   ) =  \vert 2 , 0 ),                                         \quad 
   \vert 2 E \epsilon ) =  \frac{1}{\sqrt{2}} [ \vert 2 , 2 ) + \vert 2 , -2 ) ]  \nonumber    \\
&& \vert 1 T_1 x ) 			= -\frac{i}{\sqrt{2}} [ \vert 1 , 1 ) - \vert 1 , -1 ) ]  \nonumber    \\ 
&& \vert 1 T_1 y ) 			=  \frac{1}{\sqrt{2}} [ \vert 1 , 1 ) + \vert 1 , -1 ) ]  \label{eq47} \\
&& \vert 1 T_1 z ) 			=  i \vert 1 , 0 )                                        \nonumber    \\
&& \vert 2 T_2 x ) 			=  \frac{i}{\sqrt{2}} [ \vert 2 , 1 ) + \vert 2 , -1 ) ]  \nonumber    \\
&& \vert 2 T_2 y ) 			=  \frac{1}{\sqrt{2}} [ \vert 2 , 1 ) - \vert 2 , -1 ) ]  \nonumber    \\
&& \vert 2 T_2 z ) 			= -\frac{i}{\sqrt{2}} [ \vert 2 , 2 ) - \vert 2 , -2 ) ]  \nonumber    
		\end{eqnarray}
in terms of spherical basis vectors $\vert j , m )$
(the generic symbol $\gamma$ is $a_1$ for $A_1$; $a_2$ for $A_2$; $\theta$ and $\epsilon$ for $E$; 
$x$, $y$ and $z$ for $T_1$; and $x$, $y$ and $z$ for $T_2$). The 39 independent $V$ coefficients 
are then easily calculated from (\ref{eq69}) and (\ref{eq72})-(\ref{eq47}). They 
are of course all real if we replace the pure imaginary number $i$ by $1$ in (\ref{eq47}). In 
the case $i = \sqrt{-1}$, it is possible to decrease the number of independent $V$ coefficients by 
conveniently choosing the $x(\Gamma_1 \Gamma_2 \Gamma_3)$ phase factors. Along this line, by taking 
$i = \sqrt{-1}$ and $x(\Gamma_1 \Gamma_2 \Gamma_3) = 1$ except 
$x(E T_2 T_2) = x(T_1 T_1 T_1) = x(T_1 T_1 T_2) = x(T_2 T_2 T_2) =-1$, the reader will verify that 
Eqs.~(\ref{eq69}) and (\ref{eq72})-(\ref{eq47}) lead to the real numerical values 
obtained by Griffith \cite{griffith62} for the $V$ coefficients of $O$ in his real tetragonal 
component system. 

It should be noted that each $V$ coefficient calculated from (\ref{eq69}) and (\ref{eq72})-(\ref{eq47})
can be reduced (up to a multiplicative factor) to a single 3--$jm$ coefficient for the $SO(3) \supset SO(2)$ 
chain. We thus foresee that some properties of certain 3--$jm$ symbols for the $SU(2) \supset U(1)$ chain 
may be derived by looking at some properties induced by a subgroup of $SU(2)$. As an example, we have 
 		\begin{eqnarray}
V \pmatrix{
A_2 & A_2 & E      \cr
a_2 & a_2 & \theta \cr
} \sim 
\bar f \pmatrix{
3       & 3       & 2        \cr
A_2 a_2 & A_2 a_2 & E \theta \cr
} = 
- \pmatrix{
3       & 3       & 2        \cr
-2      & 2       & 0        \cr
}
		\label{accidental}
		\end{eqnarray}
It is clear that the value of the $V$ coefficient in (\ref{accidental}) is zero since 
the $A_2 \otimes A_2 \otimes E$ triple Kronecker product does not contain the $A_1$ IRC of 
the $O$ group. As a consequence, the 3--$jm$ symbol in (\ref{accidental}) corresponding to 
the $SU(2) \supset U(1)$ chain vanishes (owing to a selection rule for $O$) in spite of the 
fact that the (trivial and Regge) symmetry properties for $SU(2) \supset U(1)$ do not impose 
such a result. 

To close Section 3, it is to be mentioned that diagrammatic methods initially developed for 
simplifying calculations within the Wigner-Racah algebra of the rotation group \cite{Yutsis}
where extended to the case of a finite or compact group \cite{Agrawala, Stedman, Elbaz}. Note 
also that considerable attention was paid in the nineties to the Wigner-Racah calculus for a 
$q$-deformed finite or compact group (see \cite{BiedLohe} for some general considerations on 
this subject and \cite{Smirnovsu(2), Asherova, Asherovasu(3)} for some developments on 
$U_q(su(2))$ and $U_q(su(3))$).

\section{Contact with quantum information}

\subsection{Computational basis and standard $SU(2)$ basis}

In quantum information, we use qubits which are nothing but state vectors in the Hilbert space 
$\mathbb{C}^2$. The more general qubit 
\begin{eqnarray}
| \psi_2 \rangle := c_0 | 0 \rangle + c_1 | 1 \rangle, 
\quad c_0 \in \mathbb{C}, 
\quad c_1 \in \mathbb{C}, 
\quad |c_0|^2 + |c_1|^2 = 1 
\end{eqnarray} 
is a linear combination of the vectors $| 0 \rangle$ and $| 1 \rangle$ which constitute an orthonormal basis
\begin{eqnarray}
B_2 := \{ | 0 \rangle, | 1 \rangle \}
\end{eqnarray}
of $\mathbb{C}^2$. The two vectors $| 0 \rangle$ and $| 1 \rangle$ can be considered as the basis vectors 
for the fundamental IRC of $SU(2)$, in the $SU(2) \supset U(1)$ scheme, corresponding to $j=1/2$ with
\begin{eqnarray}
| 0 \rangle \equiv | 1/2,  1/2 \rangle, \quad | 1 \rangle \equiv | 1/2, -1/2 \rangle
\end{eqnarray} 
More generally, in dimension $d$ we use qudits of the form
\begin{eqnarray}
| \psi_d \rangle := \sum_{n = 0}^{d-1} c_n | n \rangle,                                                      
\quad c_n \in \mathbb{C}, 
\quad n = 0, 1, \ldots, d-1, 
\quad \sum_{n = 0}^{d-1} |c_n|^2 = 1 
\end{eqnarray} 
in terms of the orthonormal basis 
\begin{eqnarray}
B_d := \{ | n \rangle : n = 0, 1, \ldots, d-1 \}
\label{Bd en n}
\end{eqnarray}
of $\mathbb{C}^d$. 
By introducing
\begin{eqnarray}
j := \frac{1}{2} (d-1), \quad m := n - \frac{1}{2} (d-1), \quad | j,m \rangle := | d-1-n \rangle 
\label{passage QI angular momentum}
\end{eqnarray}
the vectors $| n \rangle$ can be viewed as the basis vectors for the ($j$) IRC of $SU(2)$ in the 
$SU(2) \supset U(1)$ scheme. In this scheme, the $| j,m \rangle$ vector is a common eigenvector 
of the Casimir operator $J^2$ (the square of an angular momentum) and of a Cartan generator $J_z$
(the $z$ component of the angular momentum) of the $su(2)$ Lie algebra. More precisely, 
we have the relations
          \begin{eqnarray}
          J^2 |j , m \rangle = j(j+1) |j , m \rangle, \quad 
          J_z |j , m \rangle = m      |j , m \rangle
          \end{eqnarray}
which are familiar in angular momentum theory. 
In other words, the basis $B_d$, known in quantum information as the computational basis, 
can be visualized as the $SU(2) \supset U(1)$ standard basis or angular momentum basis
			\begin{eqnarray}
B_{2j+1} := \{ | j , m \rangle : m = j, j-1, \ldots, -j \}
			\end{eqnarray}
with the correspondence 
          \begin{eqnarray}
| 0   \rangle \equiv | j , j   \rangle, \quad
| 1   \rangle \equiv | j , j-1 \rangle, \quad 
\ldots, \quad
| d-1 \rangle \equiv | j , -j  \rangle
          \end{eqnarray}
between qudits and angular momentum states. 

\subsection{Nonstandard $SU(2)$ basis}
  
We are now in a position to introduce nonstandard $SU(2)$ bases which shall be connected in the next subsection 
to the so-called mutually unbiased bases (MUBs) of quantum information. As far as the representation theory of 
$SU(2)$ is concerned, we can replace the set $\{ J^2, J_z \}$ by another complete set of two commuting 
operators. Following \cite{Kib-CCCC05}, we consider the commuting set $\{ J^2, v_{ra} \}$, where the 
operator $v_{ra}$ is defined by 
          \begin{eqnarray}
          v_{ra} := {e}^{{i} 2 \pi j r} |j , -j \rangle \langle j , j| 
                  + \sum_{m = -j}^{j-1} q^{(j-m)a} |j , m+1 \rangle \langle j , m| 
          \label{definition of vra} 
          \end{eqnarray}
modulo its action on the space of constant angular momentum $j$ spanned by the $B_{2j+1}$ 
basis. In (\ref{definition of vra}), $q$ is a primitive ($2j+1$)-th root of unity, i.e.,   
          \begin{eqnarray}
          q := e^{2 \pi {i} / (2j+1)}
          \label{definition of q} 
          \end{eqnarray}
and the parameters $r$ and $a$ are fixed parameters such that
          \begin{eqnarray}       
          r \in \mathbb{R}, \quad
          a \in \mathbb{Z}/(2j+1)\mathbb{Z}
          \label{parameters} 
          \end{eqnarray}
It is to be noted that $v_{ra}$ is pseudo-invariant under the cyclic group $C_{2j+1}$ in the sense that it 
transforms as an IRC of $C_{2j+1}$ (different from the identity IRC). The common eigenstates of $J^2$ and $v_{ra}$,  
associated with the $SO(3) \supset C_{2j+1}$ chain, provide an alternative basis to that given by the common 
eigenstates of $J^2$ and $J_z$, associated with the $SO(3) \supset SO(2)$ chain. This can be precised by the 
following result. 

{\bf Theorem 3}. For fixed $j$, $r$ and $a$, the $2j+1$ common eigenvectors of $v_{ra}$ and $J^2$ can be taken in the form
          \begin{eqnarray}
|j \alpha ; r a \rangle = \frac{1}{\sqrt{2j+1}} \sum_{m = -j}^{j} 
q^{(j + m)(j - m + 1)a / 2 - j m r + (j + m)\alpha} | j , m \rangle 
          \label{j alpha r a in terms of jm}
          \end{eqnarray} 
with $\alpha = 0, 1, \ldots, 2j$. The corresponding 
eigenvalues of $v_{ra}$ are given by 
      \begin{eqnarray}
v_{ra} |j \alpha ; r a \rangle = q^{j(r+a) - \alpha} |j \alpha ; r a \rangle 
      \label{evp de vra}
      \end{eqnarray}
so that the spectrum of $v_{ra}$ is non degenerate. 

The inner product 
      \begin{eqnarray}
\langle j \alpha ; r a | j \beta ; r a \rangle = \delta_{\alpha,\beta}
      \label{jalphabetara}
      \end{eqnarray}
shows that 
      \begin{eqnarray}
B_{ra} := \{ |j \alpha ; r a \rangle : \alpha = 0, 1, \ldots, 2j \}
      \label{Bra basis}
      \end{eqnarray} 
is an orthonormal set which provides a nonstandard basis for the irreducible 
representation matrix of $SU(2)$ associated with $j$. For fixed $j$, there 
exists a ($2j+1$)-multiple infinity of orthonormal bases $B_{ra}$ since $r$ 
can have any real value and $a$, which belongs to the ring $\mathbb{Z}/(2j+1)\mathbb{Z}$, 
can take $2j+1$ distinct values ($a = 0, 1, \ldots, 2j$). 

\subsection{Other bases in quantum information}

We now go back to quantum information. By using the change of notations 
\begin{eqnarray}
d := 2j+1, \quad n := j+m, \quad | n \rangle := | j , -m \rangle, \quad |a \alpha ; r \rangle := |j \alpha ; r a \rangle
\label{passage angular momentum QI}
\end{eqnarray}
adapted to quantum information and in agreement with (\ref{passage QI angular momentum}), the operator $v_{ra}$ can be rewritten as
		\begin{eqnarray}
v_{ra} = e^{i \pi (d-1) r} \vert d-1 \rangle \langle 0 \vert + \sum_{n=1}^{d-1} q^{na} \vert n-1 \rangle \langle n \vert
		\label{vra en n bis}
		\end{eqnarray}
Each of the eigenvectors  
      \begin{eqnarray}
\vert a \alpha ; r \rangle = 
q^{(d-1)^2 r / 4}
\frac{1}{\sqrt{d}} \sum_{n = 0}^{d-1}
q^{n(d -n) a/2 + n[\alpha -(d-1)r/2]} \vert d-1-n \rangle
      \label{aalphar en n}
      \end{eqnarray}
(with $\alpha = 0, 1, \ldots, d-1$) of $v_{ra}$ is a linear combination of the qudits 
$| 0 \rangle, | 1 \rangle, \ldots, | d-1 \rangle$. For fixed $d$, $r$ and $a$, the orthonormal basis 
      \begin{eqnarray}
B_{ra} := \{ |a \alpha ; r \rangle : \alpha = 0, 1, \ldots, d-1 \}
      \label{Bra basis en aalphar}
      \end{eqnarray} 
is an alternative to the $B_d$ computational basis. As already mentioned, there is $d$-multiple infinity of orthonormal bases $B_{ra}$. 

All this can be transcribed in terms of matrices. Let ${V_{ra}}$ be the $d \times d$ matrix of the operator ${v_{ra}}$. The 
unitary matrix ${V_{ra}}$, builded on the basis $B_d$ with the ordering $0, 1, \ldots, d-1$ for the lines and columns, reads 
      \begin{eqnarray}
{{V_{ra}}} =
\pmatrix{
0                      &    q^a &      0  & \ldots &          0    \cr
0                      &      0 & q^{2a}  & \ldots &          0    \cr
\vdots                 & \vdots & \vdots  & \ldots &     \vdots    \cr
0                      &      0 &      0  & \ldots & q^{(d-1)a}    \cr
e^{  i \pi (d-1) r  }  &      0 &      0  & \ldots &          0    \cr
}
      \label{matrix Vra}
      \end{eqnarray}
The eigenvectors of ${V_{ra}}$ are 
      \begin{eqnarray}
\phi(a \alpha ; r) = q^{(d-1)^2 r / 4}
\frac{1}{\sqrt{d}} \sum_{n = 0}^{d-1}
q^{n (d-n) a/2 - n (d-1) r / 2 + n \alpha} \phi_{d-1-n}
      \label{eigenvectors of Vra}
      \end{eqnarray}
(with $\alpha = 0, 1, \ldots, d-1$), where the $\phi_{k}$ with $k = 0, 1, \ldots, d-1$ are the column vectors
      \begin{eqnarray}	
\phi_0 := \pmatrix{
1      \cr
0      \cr
\vdots \cr
0
}, \quad 
\phi_1 :=\pmatrix{
0      \cr
1      \cr
\vdots \cr
0
}, \quad 
\ldots, \quad 
\phi_{d-1}  := \pmatrix{
0      \cr
0      \cr
\vdots \cr
1
}
      \label{qudits en colonne}
      \end{eqnarray}
representing the qudits $| 0 \rangle, | 1 \rangle, \ldots, | d-1 \rangle$, respectively. They 
satisfy the eigenvalue equation 
      \begin{eqnarray}
{V_{ra}} \phi(a \alpha ; r) = q^{(d-1)(r+a)/2 - \alpha} \phi(a \alpha ; r)
      \label{eq aux valeurs propres en matrices}
      \end{eqnarray}
with $\alpha = 0, 1, \ldots, d-1$. The ${V_{ra}}$ matrix can be diagonalized by means of the ${H_{ra}}$ 
unitary matrix of elements
      \begin{eqnarray}
({H_{ra}})_{n \alpha} :=
\frac{1}{\sqrt{d}} q^{(d-1-n)(n+1) a/2 + (d-1)^2 r / 4 + (d-1-n)[\alpha -(d-1)r/2]}                         
      \label{matrix elements of Fra}
      \end{eqnarray}
with the lines and columns of ${H_{ra}}$ arranged from left to right and from 
top to bottom in the order $n, \alpha = 0, 1, \ldots, d-1$. Indeed, we have 
      \begin{eqnarray}
\left( {{H_{ra}}} \right)^{\dagger} {V_{ra}} {{H_{ra}}} =
q^{(d-1)(r+a) / 2} \pmatrix{
q^{0}                &      0 &       \ldots &          0    \cr
0                    & q^{-1} &       \ldots &          0    \cr
\vdots               & \vdots &       \ldots &     \vdots    \cr
0                    &      0 &       \ldots & q^{-(d-1)}    \cr
}
      \label{diagonalisation de Vra}
      \end{eqnarray}
in agreement with (\ref{eq aux valeurs propres en matrices}). As an illustration, let us consider 
the $d=2$ and $d=3$ cases.

For $d = 2$, we have two families of bases: the $B_{r0}$ family and the $B_{r1}$ family 
($a$ can take the values $a=0$ and $a=1$). The matrix (see (\ref{matrix Vra}))
      \begin{eqnarray}
{{V_{ra}}} :=
\pmatrix{
0                &    q^a   \cr
e^{  i \pi r  }  &      0   \cr
}, \quad q = e^{i \pi}
      \label{matrix Vra en 2 dim}
      \end{eqnarray}
has the eigenvectors (see (\ref{eigenvectors of Vra})) 
     \begin{eqnarray}
\phi ( a \alpha ; r ) = \frac{1}{\sqrt{2}} 
( q^{a/2 - r / 4 + \alpha} \phi_0 + q^{r / 4} \phi_1 ), 
\quad  \alpha = 0, 1
     \label{eigenvectors of Vra en 2 dim}
     \end{eqnarray}
which correspond to the basis $B_{ra}$. For $r=0$, the bases 
     \begin{eqnarray}
B_{00}: \quad \phi (0 0 ; 0) = \frac{1}{\sqrt{2}} \left(  \phi_1  +    \phi_0  \right), \quad 
              \phi (0 1 ; 0) = \frac{1}{\sqrt{2}} \left(  \phi_1  -    \phi_0  \right) 
     \label{dim2-1}
     \end{eqnarray}
     \begin{eqnarray}
B_{01}: \quad \phi (1 0 ; 0) = \frac{1}{\sqrt{2}} \left(  \phi_1  + i  \phi_0  \right), \quad 
              \phi (1 1 ; 0) = \frac{1}{\sqrt{2}} \left(  \phi_1  - i  \phi_0  \right) 
     \label{dim2-2}         
     \end{eqnarray}
are (up to a rearrangement) familiar bases for qubits.

For $d=3$, we have three families of bases, that is to say $B_{r0}$, $B_{r1}$ and $B_{r2}$, 
since $a$ can be 0, 1 and 2. In this case, the matrix  
      \begin{eqnarray}
{{V_{ra}}} :=
\pmatrix{
0                  &    q^a &      0  \cr
0                  &      0 & q^{2a}  \cr
e^{  i \pi 2 r  }  &      0 &      0  \cr
}, \quad q = e^{i 2 \pi / 3}
      \label{matrix Vra en dim 3}
      \end{eqnarray}
admits the eigenvectors 
      \begin{eqnarray}
\phi(a \alpha ; r) = \frac{1}{\sqrt{3}} q^{ r } 
\left( q^{a + 2 \alpha - 2 r} \phi_{0} + q^{a + \alpha - r} \phi_{1} + \phi_{2} \right), 
\quad \alpha = 0, 1, 2
      \label{eigenvectors of Vra en 3 dim}
      \end{eqnarray}
For $r=0$, the bases
     \begin{eqnarray}
B_{00}:  & & \phi(0 0 ; 0) = \frac{1}{\sqrt{3}} \left(     \phi_2 +     \phi_1 +     \phi_0 \right)  \nonumber  \\ 
         & & \phi(0 1 ; 0) = \frac{1}{\sqrt{3}} \left(     \phi_2 + q   \phi_1 + q^2 \phi_0 \right)    \\ 
         & & \phi(0 2 ; 0) = \frac{1}{\sqrt{3}} \left(     \phi_2 + q^2 \phi_1 + q   \phi_0 \right)  \nonumber  \\
B_{01}:  & & \phi(1 0 ; 0) = \frac{1}{\sqrt{3}} \left(     \phi_2 + q   \phi_1 + q   \phi_0 \right)  \nonumber  \\ 
         & & \phi(1 1 ; 0) = \frac{1}{\sqrt{3}} \left(     \phi_2 + q^2 \phi_1 +     \phi_0 \right)    \\ 
         & & \phi(1 2 ; 0) = \frac{1}{\sqrt{3}} \left(     \phi_2 +     \phi_1 + q^2 \phi_0 \right)  \nonumber  \\
B_{02}:  & & \phi(2 0 ; 0) = \frac{1}{\sqrt{3}} \left(     \phi_2 + q^2 \phi_1 + q^2 \phi_0 \right)  \nonumber  \\
         & & \phi(2 1 ; 0) = \frac{1}{\sqrt{3}} \left(     \phi_2 +     \phi_1 + q   \phi_0 \right)    \\ 
         & & \phi(2 2 ; 0) = \frac{1}{\sqrt{3}} \left(     \phi_2 + q   \phi_1 +     \phi_0 \right)  \nonumber          
     \end{eqnarray}
are useful for qutrits. 

\subsection{Mutually unbiased bases}
Going back to the case where $d$ is arbitrary, we now examine an important property of the couple ($B_{ra}, B_d$) and its generalization 
to couples ($B_{ra}, B_{rb}$) with $b \not= a$. For fixed $d$, $r$ and $a$, (\ref{aalphar en n}) gives 
      \begin{eqnarray}
\forall n, \alpha \in \{ 0, 1, \ldots, d-1 \} : 
\vert \langle n | a \alpha ; r \rangle \vert = \frac{1}{\sqrt{d}} 
      \label{Bra et Bd}
      \end{eqnarray}
Equation (\ref{Bra et Bd}) shows that $B_{ra}$ and $B_d$ are two unbiased bases. (Let us recall that 
two distinct orthonormal bases $B_a = \{ | a \alpha \rangle : \alpha = 0, 1, \ldots, d-1 \}$ and 
                      $B_b = \{ | b \beta  \rangle : \beta  = 0, 1, \ldots, d-1 \}$ of the Hilbert space 
$\mathbb{C}^{d}$ are said to be unbiased if and only if the inner product 
$\langle a \alpha | b \beta \rangle$ has a modulus independent of $\alpha$ and 
$\beta$.) 

Other examples of unbiased bases can be obtained for $d = 2$ and $3$. We easily verify that the bases $B_{r0}$ and 
$B_{r1}$ for $d=2$ given by (\ref{eigenvectors of Vra en 2 dim}) are unbiased. Similarly, the bases $B_{r0}$, $B_{r1}$ 
and $B_{r2}$ for $d=3$ given by (\ref{eigenvectors of Vra en 3 dim}) are mutually unbiased. Therefore, by combining 
these particular results with the general result implied by (\ref{Bra et Bd}) we end up with 3 mutually unbiased bases 
(MUBs) for $d=2$ and 4 MUBs for $d=3$. This is in agreement with the theorem according to which the number 
$N_{\scriptscriptstyle MUB}$ of pairwise MUBs in $\mathbb{C}^d$ is such that $3 \leq N_{\scriptscriptstyle MUB} \leq d+1$ 
and that the maximum number $d+1$ is attained when $d$ is a prime number $p$ or an integer power $p^e$ ($e \geq 2$) of a 
prime number $p$ \cite{Ivanovic,Wootters,Calderbank}. The results for $d=2$ and $3$ can be 
generalized in the case where $d$ is a prime number. This can be precised by the following 
theorem \cite{KibIJMPB2006, KibPla, AlbKib, Kib2008, Kibler09}.  

{\bf Theorem 4}. For $d=p$, with $p$ a prime number, the bases $B_{r0}, B_{r1}, \ldots, B_{rp-1}, B_{p}$ 
corresponding to a fixed value of $r$ form a complete set of $p+1$ MUBs. The $p^2$ vectors 
$| a \alpha ; r \rangle$, with $a, \alpha = 0, 1, \ldots, p-1$, of the bases  
$B_{r0}, B_{r1}, \ldots, B_{rp-1}$ are given by a single formula, namely (\ref{aalphar en n}) 
or (\ref{eigenvectors of Vra}). The index $r$ makes it possible to distinguish different complete 
sets of $p+1$ MUBs.

The proof is as follows. First, according to (\ref{Bra et Bd}), the computational basis $B_{p}$ is 
unbiased with any of the $p$ bases $B_{r0}, B_{r1}, \ldots, B_{rp-1}$. Second, we get 
    \begin{eqnarray}	                  
\langle a \alpha ; r | b \beta ; r \rangle = \frac{1}{p}     
\sum_{k = 0}^{p-1} q^{k(p-k)(b-a) / 2 + k(\beta - \alpha)} 
    \label{produit scalaire} 
	  \end{eqnarray} 
or 
     \begin{eqnarray}
\langle a \alpha ; r | b \beta ; r \rangle = \frac{1}{p} \sum_{k = 0}^{p-1} 
e^{i \pi \{ (a-b)k^2 + [p(b-a) + 2(\beta - \alpha)]k \} / p}
     \label{inner product p prime}
     \end{eqnarray}
The right-hand side of (\ref{inner product p prime}) can be expressed 
in terms of a generalized quadratic Gauss sum \cite{BerndtEW} 
     \begin{eqnarray}
     S(u, v, w) := \sum_{k = 0}^{|w|-1} e^{i \pi (u k^2 + v k) / w}
     \label{Suvw}
     \end{eqnarray}
where $u$, $v$ and $w$ are integers such that $u$ and $w$ are mutually prime, 
$uw$ is non vanishing and $uw + v$ is even. This leads to 
     \begin{eqnarray}
\langle a \alpha ; r | b \beta ; r \rangle = \frac{1}{p} S(u, v, w)   
     \label{G1}
     \end{eqnarray}
with 
     \begin{eqnarray}
     u := a - b, \quad v := -(a - b)p - 2(\alpha - \beta), \quad w := p
     \label{G2}
     \end{eqnarray}
The generalized Gauss sum $S(u, v, w)$ in (\ref{G1})-(\ref{G2}) can be calculated from the 
methods described in \cite{BerndtEW}. We thus obtain 
     \begin{eqnarray}
 | \langle a \alpha ; r | b \beta ; r \rangle | = \frac{1}{\sqrt{p}} 
     \label{module du ps BraBrb}
     \end{eqnarray}
which completes the proof.$\Box$

\subsection{Mutually unbiased bases and Lie algebras}

\subsubsection{Weyl pairs}

The matrix ${V_{ra}}$ can be decomposed as 
     \begin{eqnarray}
{V_{ra}} = {P_r} {X} {Z}^a
     \end{eqnarray}
where
       \begin{eqnarray}
{P_r} := 
\pmatrix{
1                    &      0 &      0    & \ldots &       0                 \cr
0                    &      1 &      0    & \ldots &       0                 \cr
0                    &      0 &      1    & \ldots &       0                 \cr
\vdots               & \vdots & \vdots    & \ldots &       \vdots            \cr
0                    &      0 &      0    & \ldots &       e^{i \pi (d-1) r} \cr
}
        \end{eqnarray}
and
        \begin{eqnarray}
{X} := 
\pmatrix{
0                    &      1 &      0  & \ldots &       0 \cr
0                    &      0 &      1  & \ldots &       0 \cr
\vdots               & \vdots & \vdots  & \ldots &  \vdots \cr
0                    &      0 &      0  & \ldots &       1 \cr
1                    &      0 &      0  & \ldots &       0 \cr
}, \quad 
{Z} := 
\pmatrix{
1                    &      0 &      0    & \ldots &       0       \cr
0                    &      q &      0    & \ldots &       0       \cr
0                    &      0 &      q^2  & \ldots &       0       \cr
\vdots               & \vdots & \vdots    & \ldots &  \vdots       \cr
0                    &      0 &      0    & \ldots &       q^{d-1} \cr
}
        \end{eqnarray}
The linear operators corresponding to the matrices ${X}$ and ${Z}$ are known in quantum information 
as shift and clock operators, respectively. The unitary matrices ${X}$ and ${Z}$ $q$-commute in the 
sense that 
     \begin{eqnarray}
{X} {Z} - q {Z} {X} = 0 
     \label{qcom of Xr and Z}
     \end{eqnarray}
In addition, they satisfy
     \begin{eqnarray}
{X}^d = {Z}^d = {I_d} 
     \label{dpower of Xr and Z}
     \end{eqnarray}
where ${I_d}$ is the $d$-dimensional unit matrix. Equations (\ref{qcom of Xr and Z}) and 
(\ref{dpower of Xr and Z}) show that ${X}$ and ${Z}$ constitute a Weyl pair \cite{Weyl1931}. The 
(${X} , {Z}$) Weyl pair turns out to be an integrity basis for generating a set 
$\{ {X}^a {Z}^b : a,b = 0, 1, \ldots, d-1 \}$ of $d^2$ generalized Pauli matrices in $d$ 
dimensions (see for instance \cite{Kib2008, BandyoGPM5, LawrenceGPM6, PittengerRubin} in the context 
of MUBs and \cite{Tolar1,Balian,PateraZassenhaus} in group-theoretical contexts). In addition, 
the set $\{ q^a {X}^b {Z}^c : a,b,c = 0, 1, \ldots, d-1 \}$ 
generates, with respect to matrix multiplication, a finite group of order $d^3$, the $P_d$ Pauli group 
\cite{Kib2008}. As an example, for $d=2$ we have 
\begin{eqnarray}
{X}      =     \sigma_x, \quad   
{Z}      =     \sigma_z, \quad 
{X}{Z}     = - i \sigma_y, \quad 
{X}^0{Z}^0 =     \sigma_0
\end{eqnarray}
in terms of the ordinary Pauli matrices $\sigma_0 = I_2$, $\sigma_x$, $\sigma_y$ and $\sigma_z$, and the 
Pauli group $P_2$ is isomorphic with the hyperbolic quaternion group. 

Equations (\ref{qcom of Xr and Z}) and (\ref{dpower of Xr and Z}) can be generalized through 
     \begin{eqnarray}
{V_{ra}} {Z} - q {Z} {V_{ra}} = 0, \quad e^{-i \pi (d-1)(r+a)} ({V_{ra}})^d = {Z}^d = {I_d}
     \end{eqnarray}
so that other pairs of Weyl can be obtained from ${V_{ra}}$ and ${Z}$. Note that
     \begin{eqnarray}
{X} = {V_{00}}, \quad 
{Z} = \left( {V_{00}} \right)^{\dagger} {V_{01}}
     \end{eqnarray} 
which shows a further interest of the matrix ${V_{ra}}$.

\subsubsection{MUBs and the special linear group}

In the case where $d$ is a prime integer or a power of a prime integer, it is known that the set 
$\{ {X}^a{Z}^b : a, b = 0, 1, \ldots, d-1 \} \setminus \{ {X}^0{Z}^0 \}$ of 
cardinality $d^2 - 1$ can be partitioned into $d+1$ subsets containing each $d-1$ commuting 
matrices (cf.~\cite{BandyoGPM5}). Let us give an example.

For $d=5$, we have the 6 following sets of 4 commuting matrices
           \begin{eqnarray}  	   
{\cal V}_0      &:=      &  \{  01 ,  02 ,  03 ,  04  \}, 
  \quad	   	   
{\cal V}_1       :=         \{  10 ,  20 ,  30 ,  40  \} 
  \nonumber \\
{\cal V}_2      &:=      &  \{  11 ,  22 ,  33 ,  44  \}, 
  \quad
{\cal V}_3       :=         \{  12 ,  24 ,  31 ,  43  \} 
  \label{eq152bis} \\
{\cal V}_4      &:=      &  \{  13 ,  21 ,  34 ,  42  \}, 
  \quad
{\cal V}_5       :=         \{  14 ,  23 ,  32 ,  41  \} 
  \nonumber
          \end{eqnarray} 
where $ab$ is used as an abbreviation of ${X}^a {Z}^b$.

More generally, for $d=p$ with $p$ prime, the $p+1$ sets of $p-1$ commuting matrices 
are easily seen to be                 
           \begin{eqnarray}  
{\cal V}_0       &:=      &  \{ {X}^0 {Z}^a        :  a = 1, 2, \ldots, p-1 \} 
    \nonumber \\                   
{\cal V}_1       &:=      &  \{ {X}^a {Z}^0        :  a = 1, 2, \ldots, p-1 \}   
    \nonumber \\
{\cal V}_2       &:=      &  \{ {X}^a {Z}^a        :  a = 1, 2, \ldots, p-1 \} 
    \nonumber \\
{\cal V}_3       &:=      &  \{ {X}^a {Z}^{2a}     :  a = 1, 2, \ldots, p-1 \} 
              \\
                 &\vdots  & 
    \nonumber \\
 {\cal V}_{p-1}  &:=      &  \{ {X}^a {Z}^{(p-2)a} :  a = 1, 2, \ldots, p-1 \} 
    \nonumber \\  
 {\cal V}_{p}    &:=      &  \{ {X}^a {Z}^{(p-1)a} :  a = 1, 2, \ldots, p-1 \} 
    \nonumber 
           \end{eqnarray} 
Each of the $p+1$ sets ${\cal V}_0, {\cal V}_1, \ldots, {\cal V}_{p}$ can be put in a one-to-one correspondence 
with one basis of the complete set of $p+1$ MUBs. In fact, ${\cal V}_0$ is associated with the computational basis 
while ${\cal V}_1, {\cal V}_2, \ldots, {\cal V}_{p}$ are associated with the $p$ remaining MUBs in view of 
           \begin{eqnarray} 
{V_{0 a}} \in {\cal V}_{a + 1} = \{ {X}^b {Z}^{ab} : b = 1, 2, \ldots, p-1 \}, \quad a = 0, 1, \ldots, p-1 
           \end{eqnarray} 
Keeping into account the fact that the set $\{ {X}^a {Z}^b : a,b = 0, 1, \ldots, p-1 \} \setminus \{ {X}^0 {Z}^0 \}$
spans the Lie algebra of the special linear group $SL(p, \mathbb{C})$, we have the following theorem.

{\bf Theorem 5}. For $d=p$, with $p$ a prime integer, the Lie algebra 
$sl(p, \mathbb{C})$ of the group $SL(p, \mathbb{C})$ can be decomposed into a sum 
(vector space sum) of $p+1$ abelian subalgebras each of dimension $p-1$, i.e.
                  \begin{eqnarray}
sl(p, \mathbb{C}) \simeq 
{ v}_0     \uplus 
{ v}_1     \uplus 
\ldots     \uplus      
{ v}_{p}     
                  \end{eqnarray}
where the $p+1$ subalgebras ${ v}_0, { v}_1, \ldots, { v}_p$ are 
Cartan subalgebras generated respectively by the sets ${\cal V}_0, {\cal V}_1, \ldots, {\cal V}_{p}$ 
containing each $p - 1$ commuting matrices.

The latter result can be extended when $d = p^e$ with $p$ a prime integer and 
$e$ an integer ($e \geq 2$): there exists a decomposition of $sl(p^e, \mathbb{C})$ into $p^e + 1$ 
abelian subalgebras of dimension $p^e - 1$ (cf.~\cite{Kibler09, PateraZassenhaus, autresdecomp3, autresdecomp2}). 

\section{Appendix: The Racah parameters} 

In the case of the $\ell^N$ configuration, the Coulomb Hamiltonian ${\cal H}_C$ can be written as 
                  \begin{eqnarray}
{\cal H}_C = (2 \ell + 1)^2 \sum_{k = 0, 2, \ldots, 2 \ell} F^{k} 
\left ( 
{\matrix {\ell & k & \ell \cr 0&0&0}} 
\right )^2 
\sum_{i<j} 
\left( 
{\bf u}^{(k)}(i) \cdot {\bf u}^{(k)}(j) 
\right)
                  \end{eqnarray}
where the $F^{k} \equiv D_k(\ell) F_k$ parameters  are the usual Slater-Condon-Shortley parameters. It is 
clear that any linear transformation
		\begin{eqnarray}
{\cal E}^\lambda = \sum_{k = 0, 2, \ldots, 2 \ell} b(\ell)_{\lambda k} F^{k}, \quad \lambda = 0, 1, \ldots, \ell 
		\label{def des Elambda}
		\end{eqnarray}
where $b(\ell)$ is a regular matrix of dimension $\ell + 1$ defines an equally acceptable parametrization. 

As a trivial example, 
the $D[\ldots]$ parametrization in Section 2.5 corresponds to 
		\begin{eqnarray}
D[(00)0(kk)00] = (s \Vert u^{(0)} \Vert s)^{-2} (\ell \Vert u^{(k)} \Vert \ell)^{-2} 
\sqrt{2k+1} (2 \ell + 1)^2 \left ( {\matrix {\ell & k & \ell \cr 0&0&0}} 
\right )^2 F^{k} 
		\end{eqnarray}
i.e., to a renormalization of the $F^{k}$ parameters. 

Less trivial examples are provided by the Racah parameters 
		\begin{eqnarray}
&& A = F^{0} - \frac{1}{9} F^{4} = F_0 - 49 F_4 \nonumber \\		
&& B = \frac{1}{441} ( 9 F^{2} - 5 F^{4} ) = F_2 - 5 F_4 \label{ABC} \\		
&& C = \frac{5}{63} F^{4} = 35 F_4 \nonumber
		\end{eqnarray}
for the $d^N$ configuration \cite{RII} and the Racah parameters 
		\begin{eqnarray}
&& E^0 = F_0 - 10 F_2 - 33 F_4 - 286 F_6 \nonumber \\		
&& E^1 = \frac{1}{9} (70 F_2 + 231 F_4 + 2002 F_6) \nonumber \\	 	
&& E^2 = \frac{1}{9} (F_2 - 3 F_4 + 7 F_6) \label{E0E1E2E3} \\		
&& E^3 = \frac{1}{3} (5 F_2 + 6 F_4 - 91 F_6) \nonumber \\
&& F_0 = F^{0}, \quad F_2 = \frac{1}{225} F^{2}, \quad F_4 = \frac{1}{1089} F^{4}, \quad F_6 = \frac{25}{184081} F^{6} \nonumber
		\end{eqnarray}
for the $f^N$ configuration \cite{RIV}. The term energies for $d^N$ assume, to some extent, a simple form when expressed as 
functions of $A$, $B$ and $C$. The $E^j$ parameters (with $j = 0, 1, 2, 3$) for $f^N$ allow to decompose ${\cal H}_C$ into 
parts having well-defined properties under the action of the groups of the $SO(7) \supset G_2 \supset SO(3)$ chain. 

As a last example, let us consider the parametrization defined by (\ref{def des Elambda}) with 
		\begin{eqnarray}
b(\ell)_{\lambda k} = (-1)^\lambda 
(2 \ell + 1) 
\left ( 
{\matrix {\ell & k & \ell \cr 0&0&0}} 
\right ) 
\left ( 
{\matrix {\ell & k & \ell \cr -\lambda&0&\lambda}} 
\right )
		\end{eqnarray}
In this parametrization, the ${\cal H}_C$ operator can be rewritten as  
		\begin{eqnarray}
{\cal H}_C = \sum_{\lambda = -l}^{l} V_{\lambda} 
		\end{eqnarray}
with
		\begin{eqnarray}
V_{\lambda} = {\cal E}^{\lambda} \sum_{k = 0, 2, \ldots, 2 \ell} (2k + 1) b(\ell)_{\lambda k}
                                 \sum_{i<j}
\left(
{\bf u}^{(k)}(i) \cdot {\bf u}^{(k)}(j)
\right)
		\end{eqnarray}
We of course have $V_{\lambda} = V_{-\lambda}$ and therefore there are $\ell+1$ independent components $V_{\lambda}$ 
in ${\cal H}_C$. The ${\cal E}^{\lambda}$ parametrization was investigated in \cite{alternative, KibKat, KibPar}. Let 
us simply mention that the part $V_0$ of ${\cal H}_C$ corresponds to a sum of surface delta interactions and that 
${\cal H}_C$ can be reduced to $V_0$ for 
		\begin{eqnarray}
F^{k} = (2 k + 1) F^{0}
		\label{B2B4}
		\end{eqnarray}
for $k = 0, 2, \ldots, 2 \ell$. In the special case of the $d^N$ configuration, it is to be 
realized that relation (\ref{B2B4}) corresponds to the Laporte-Platt degeneracies \cite{Laporte} 
(see also \cite{Juddbook, alternative, King}) which occur for $B=0$. 

\section{Closing remarks}

Starting with the idea to substitute for the numerical methods of Slater, Condon and 
Shortley general methods close both to Dirac's ideas on quantum mechanics 
and to those of Wigner about the use of symmetries in physics, Racah developed practically 
in 20 years universal methods (irreducible tensor methods and group theoretical methods) 
used in many fields of physics and chemistry. In particular, the application of Racah's 
methods in atomic, nuclear and elementary particle physics as well as in group theory 
(Wigner-Racah algebra, state labeling problem) are well-known. We have shown how the use 
of Racah's methods in conjunction with $SU(2) \supset G^*$ or $SO(3) \supset G$ symmetry 
adapted bases and effective operators yields sophisticated models in crystal- and 
ligand-field theories. In last analysis, these models are fully described by chains 
of groups, viz., the $U(5) \supset SO(5) \supset SO(3) \supset G$ chain 
for the $d^N$ configuration in $G$ and the $U(7) \supset SO(7) \supset G_2 \supset SO(3) \supset G$ 
chain for the $f^N$ configuration in $G$.  

As an application of current interest in the present days, we have shown the importance of the 
chain $SO(3) \supset C_d$ for deriving a complete set of mutually unbiased bases when $d$ is a 
prime integer. These bases are very useful in quantum information (quantum cryptography, 
quantum state tomography, quantum error codes) and equally in quantum mechanics 
(discrete Wigner function, mean King problem, path integral formalism). 

A common denominator to Sections 2, 3 and 4 is the notion of ``chains of groups''. Although chains of groups were 
in use before Racah (e.g., see the works by Ehlert on CH$_4$ \cite{CH4}, Bethe on crystal-field theory 
\cite{Bethe} and Wigner on supermultiplets of nuclei \cite{multiplets}), his contribution to that part 
of applied group theory is essential and represents one of its major achievements.\footnote{A 
fundamental result proved by Racah is that for a chain of groups having for head group a Lie group of 
order $r$ and rank $l$, one can associate a complete set of commuting operators of cardinal 
$(r + l)/2$ (i.e., $l$ Cartan operators plus $l$ Casimir operators plus $(r - 3l)/2$ labeling 
operators, some of the operators being Casimir, Cartan or labeling operators of the chain) 
\cite{IASP}. See \cite{Boya} for recent developments on this subject.} The 
interest for Physics and Chemistry of chains involving (noncompact and/or compact) continuous as 
well as finite groups is now well established. Such chains turn out to be useful in the investigation 
of broken symmetries which may arise either via descent in symmetry (Zeeman effect, homogeneous and 
inhomogeneous Stark effect, ligand-field effect, etc.) or via spontaneous symmetry breaking (Landau 
and Jahn-Teller effect, symmetry breaking in elementary particle physics, etc.). In Racah's approach, 
which excludes the cases of external or Lorentzian and internal or gauge (super)symmetries, one group 
of the chain is a high symmetry group corresponding to a zeroth order approximation (like the cubic 
group in ligand-field theory) and another one is a low symmetry group corresponding to a first order 
approximation (like the tetragonal or trigonal group in ligand-field theory). The two symmetry groups 
correspond to known or postulated symmetries depending on whether the nature of the interactions involved
is known or unknown. According to Wigner's theorem \cite{wigner27}, these symmetry or invariance groups 
(which leave invariant an Hamiltonian operator) provide representation labels or good quantum numbers for 
describing the state vectors. The other groups of the chain are dynamical or noninvariance groups 
in the sense that not all of its generators or elements commute with the Hamiltonian. They 
can describe part of the interactions and are generally introduced to make the chain as multiplicity-free 
as possible. Finally, the various groups of the chain are used to classify the state vectors and the 
(known or postulated) physical interactions. When elaborating a model based on symmetry considerations, 
the latter point is of considerable importance from a qualitative point of view (for level splitting and 
for selection rules) and a quantitative point of view (for the calculation of energy or mass matrices 
and transition probabilities). The preceding considerations apply to nuclear, atomic, molecular and 
condensed matter physics and also to quantum chemistry (chains of groups are even useful for classifying 
chemical elements \cite{Bar72-27, RumFet72-28, KibMolPhys}). Note that the situation is a bit different 
in elementary particle physics since the notion of classification groups (with the pioneer works by 
Heisenberg, Sakata, Gell-Mann Ne'eman and Zweig going from the $SU(2)$ isospin group to the 
$SU(3) \supset SU(2)$ chain involved in the first quark model) evolved to gauge groups (going from the 
$SU(3) \otimes SU(2) \otimes U(1) \supset SU(3) \otimes U(1)$ standard model to the grand unified 
models based on the $E_8 \supset E_7 \supset E_6 \supset SO(10) \supset SU(5) \supset 
SU(3) \otimes SU(2) \otimes U(1) \supset SU(3) \otimes U(1)$). However, in any field of physics there is 
a common scheme, namely, \\ 
\begin{center}
(super)symmetries $\to$ chain of groups $\to$ invariance or \\ 
co-variance $\to$ conservation laws or good quantum numbers. 
\end{center}

To close this paper, let us add some further comments. Racah founded the main school of theoretical physics 
in Israel. He had a strong impact on (national and international) committees and on various research groups 
in theoretical and experimental spectroscopy (including the Laboratoire Aim\'e Cotton in France). 

Racah had many students who deeply contributed to atomic and nuclear spectroscopy; they 
are profusely quoted in the review by Zeldes \cite{Zeldes}. We would like to complete the list 
of students in the bibliography of Zeldes with a few words about Mosh\'e Flato (Tel Aviv 
1937 - Paris 1998), a student of Racah during the period 1959-1963, who contributed to spread 
the ideas of Racah on crystal- and ligand-field theories. Flato achieved his M.Sc. thesis under 
the supervision of Racah in 1959 and prepared in 1960-1963 a Ph.D. thesis on a subject of nuclear 
physics (dealing with the $Sp(2n) \supset U(n)$ chain in connection with the Elliott-Flowers model) 
given by Racah.\footnote{According to Daniel 
Sternheimer, Flato was Racah's preferred student, probably the most brilliant in his generation. The families were 
friends since WWII when Flato's father was chief engineer of the British Mandate in Jerusalem. When Racah became 
Rector, he asked Flato to deliver (during 2 years, while Flato was doing military service) the traditional 
Racah lectures on group theory in physics, and recommended him for a course on solid state physics at Bar 
Ilan University, which Flato delivered without the compulsory yarmulke to students about his age.} When Flato 
came to France in 1963 his interest shifted to noncompact groups. He started working on Lorentzian symmetries 
and strong interactions when he was at Institute Henri Poincar\'e in Paris (1963-1964). After that he was 
Associate Professor of Physics for three years at Universit\'e de Lyon, then moved to Dijon in 
mathematics. Flato got in 1965 a Doctorat \`es Sciences Physiques from Universit\'e de Paris
on the basis of his works on elementary particle physics.\footnote{According to Sternheimer, after the death of Racah 
in 1965 (in Firenze on his way to join Flato in Paris) Flato decided not to publish the joint paper which they 
were preparing in nuclear physics. Racah had taken the manuscript with him in Firenze and intended to finalize 
it in Paris. Flato did not either defend in Jerusalem his Ph.D. based on that paper. Anyhow that became moot since 
he had already a French D.Sc.} Flato pursued a brilliant career both in France and worldwide, dealing with 
a great variety of subjects in physics and in mathematics. Flato evolved from theoretical 
physics to mathematical physics and mathematics.\footnote{According to Sternheimer, his coworker for 35 years who 
heard with him a course on the theory of distributions by S. Agmon in Jerusalem in 1958-59, Flato had a dual 
training in physics and mathematics. Before opting for Racah he had considered working with S. Amitsur in algebra 
or with N. Rosen in relativity.} Among his many interests and contributions, let us mention 
the following: mass formulas (in relation with internal and external symmetries), 
conformal field theories; infinite-dimensional representations of Lie groups, singletons, 
AdS$_4$/CFT$_3$, composite electrodynamics; nonlinear representations of groups, covariant 
PDEs, global existence theorems for field theories (Yang-Mills, Maxwell-Dirac); and  
especially the role of deformations in physics, including the now 35 years old and 
still frontier area of deformation quantization (symplectic and Poisson manifolds), 
quantum groups and noncommutative geometry. In 1968 Flato founded what was later called the 
Laboratoire Gevrey de Math\'ematique Physique at Universit\'e de Bourgogne and in 1975 Letters in 
Mathematical Physics and two series of books published by Reidel (Kluwer). He had numerous 
students in France and abroad and a strong impact on Society (IAMP, Marie Curie chairs, Scientific 
Council of UAP). For more details, see \cite{http1, http2}. 

Racah and Flato shared important scientific and human qualities. Both were 
excellent teachers and at the same time exceptional researchers with a good sense of the duality 
theory-experiment, convinced of the importance of symmetries in physics. They knew how 
to communicate enthusiasm, give the right impulse to their students and collaborators, and 
inspire them to solve problems. Both had a strong impact on scientific communities and 
on national and international committees and enterprises. We learned and can still learn many
things from them both from the human and scientific points of view. Their impact will last for a 
long time. We shall not forget them. 

{\bf Acknowledgements}
\\
This paper is based on an invited talk given at the ``International Conference in Commemoration 
of the Centenary of the Birth of G. Racah (1909-1965)'' (Zaragoza, Spain, 22-24 February 2010) 
organized by the Real Academia de Ciencias de Zaragoza and sponsored by the Fundaci\'on Ram\'on 
Areces. The author thanks the organizers L.J. Boya and R. Campoamor-Stursberg for having made 
this beautiful conference in honor of Racah possible. He also thank F. Bartolom\'e, L.J. Boya, 
R. Campoamor-Stursberg, H. de Guise, G. Delgado-Barrio, M.A. del Olmo, S. Elitzur, F. Iachello, M. Santander, 
I. Unna, and P. Van Isacker for interesting discussions. 

My indebted thoughts go to my friend and teacher M. Flato who introduced me to Racah's methods, 
supervised my thesis with continuous guidance and gave me the taste of theoretical physics. 

Many other people contributed to my knowledge of Racah's methods and their application to the physics of ions 
in crystals. I would like to thank: C.K. J{\o}rgensen, Tang Au-chin, J. Patera, P. Winternitz, Y. Ilamed, 
and B.G. Wybourne for fruitful correspondence and/or discussions; Y. Bordarier who helped me to use 
programs set up in Racah's laboratory; some of my students: P. Guichon, G. Grenet, T. N\'egadi, 
M. Daoud, and C. Campigotto; some of my colleagues and collaborators: F. Gaume, C. Linar\`es, A. Louat, 
J.-C. G\^acon, J.C. Souillat, A. Gros, B. Jacquier, M. Bouazaoui, J.F. Marcerou, Y. Jugnet, Tran Minh Duc, 
J. Katriel, J. Sztucki and Yu.F. Smirnov, and R.M. Asherova. 

I would also like to acknowledge my student O. Albouy and my colleagues M. Planat and M. Saniga for recent 
collaboration on quantum information. 

Finally, thanks are due to D. Sternheimer for numerous discussions and 
correspondence, and his help in the oral and written presentation of 
this paper. 



\end{document}